\documentclass[aps,twocolumn]{revtex4}
\usepackage{amsmath,amssymb,amsthm,amsfonts}
\usepackage{lineno,url}\usepackage{float}
\floatstyle{plaintop}\restylefloat{table}
\usepackage[tableposition=top]{caption}
\usepackage{graphicx,epsfig}
\usepackage{multirow}\usepackage{pstricks}\usepackage{fancyvrb}
\usepackage{listings,paralist} 
\newtheorem{theorem}{Theorem}[section]
\theoremstyle{definition}
\newtheorem{remark}[theorem]{Remark}
\newlength{\tew}

\def\ig{\includegraphics}

\def\medskip{}\def\bigskip{}

\def\twi{\textwidth}
\def\uh{u_{{\rm hom}}}\def\uhs{u_{{\rm hom}}^{{\rm small}}}
\def\uhl{u_{{\rm hom}}^{{\rm large}}}
\def\up{u_{{\rm p},1}}\def\ups{u_{{\rm p},1}^{{\rm small}}}
\def\upl{u_{{\rm p},1}^{{\rm large}}}\def\upsg{u_{{\rm p}}^{{\rm small}}}
\def\uplg{u_{{\rm p}}^{{\rm large}}}\def\um{u^{\rm middle}}
\def\hecy{heteroclinic cycle}
\def\pdep{{\tt pde2path}}
\newcommand{\R}{{\mathbb R}}\newcommand{\N}{{\mathbb N}}
\newcommand{\bci}{\begin{compactitem}}\newcommand{\eci}{\end{compactitem}}
\newcommand{\bcen}{\begin{compactenum}}\newcommand{\ecen}{\end{compactenum}}
\newcommand{\hs}[1]{{\hspace{#1}}}\newcommand{\vs}[1]{{\vspace{#1}}}
\def\dd{\, {\rm d}}\def\CE{{\cal E}}
\def\brem{\begin{remark}}\def\erem{\end{remark}}
\def\eex{\hfill\mbox{$\rfloor$}}
\def\noi{\noindent}
\newcommand{\bce}{\begin{center}}\newcommand{\ece}{\end{center}}
\newcommand{\reff}[1]{(\ref{#1})}
\def\huga#1{\begin{gather} #1 \end{gather}}

\def\pa{{\partial}}\def\lam{\lambda}\def\Om{\Omega}
\def\ee{{\rm e}}\def\ii{{\rm i}}

\usepackage{hyperref}

\begin{document}
\title{Defect-like structures and localized patterns in SH357}
\author{Edgar Knobloch$^1$, Hannes Uecker$^{2}$\thanks{hannes.uecker@uni-oldenburg.de}, 
  Daniel Wetzel$^2$}
\affiliation{\vspace{3mm}$^1$Department of Physics, University of California, Berkeley, CA 94720, USA\\
$^2$Institut f\"ur Mathematik, Universit\"at Oldenburg, D26111 Oldenburg, 
Germany}

\begin{abstract} We study numerically the cubic-quintic-septic Swift-Hohenberg (SH357) equation on bounded one-dimensional domains. Under appropriate conditions stripes with wave number $k\approx 1$ bifurcate supercritically from the zero state and form S-shaped branches resulting in bistability between small and large amplitude stripes. Within this bistability range we find stationary heteroclinic connections or fronts between small and large amplitude stripes, and demonstrate that the associated spatially localized defect-like structures either snake or fall on isolas. In other parameter regimes we also find heteroclinic connections to spatially homogeneous states, and a multitude of dynamically stable steady states consisting of patches of small and large amplitude stripes with different wave numbers or of spatially homogeneous patches. The SH357 equation is thus extremely rich in the types of patterns it exhibits. Some of the features of the bifurcation diagrams obtained 
 by numerical continuation can be understood using a conserved quantity, the spatial Hamiltonian of the system.
\end{abstract}

\maketitle

\section{Introduction}\label{isec}

Defects play an important role in both condensed matter physics and in pattern-forming systems. Topological defects arise when the pattern amplitude vanishes and organize both one- and two-dimensional patterns \cite{ch93,pismen06}. As a result their motion and creation or annihilation lead to nonlocal adjustment of the pattern wave number. Nontopological defects may involve a continuous but spatially localized transition from one wave number to another, and such transitions are usually referred to as fronts \cite{ch93,pismen06}. Fronts connecting distinct states are common in systems supporting waves as exemplified by the complex Ginzburg-Landau equation \cite{huber,tpk,motter} and pattern formation in fluid dynamics \cite{spina} and chemistry \cite{epstein}, and are found in both wave-supporting systems and in stationary patterns. A mathematical classification of defects in one spatial dimension is provided in Ref.~\cite{scheel}.

In this paper we are interested in properties of stationary fronts between patterns with distinct wave numbers in one spatial dimension -- the simplest defect -- and ask two basic questions. Do such defects come in distinct families and do they persist over time? The first question is motivated by the notion of pinning \cite{pomeau} whereby a front is pinned to heterogeneities on either side of the front which can be viewed as trapping the front in a particular location. Energy has to be supplied to overcome the resulting pinning potential to move the defect to a different location. The second question has to do with stability of a front. A front may lose stability because of the loss of stability of either of the far-field states or due to a localized mode at the location of the front. These instabilities are usually associated with the presence of unstable continuous and point spectra, respectively. We remark that the classical amplitude-phase description of patterns misses
  the pinning effect just described since it reduces a uniform pattern to a constant amplitude state and hence treats such states as translation-invariant states. To study fronts between different wave numbers it is essential therefore to go beyond the amplitude-phase description. 

The fact that defects are spatially localized structures and that such
structures readily pin to heterogeneities suggests that they might
come in families organized within a snaking bifurcation diagram. In
the context of pattern-forming systems the term homoclinic snaking
refers to a pair of intertwined branches of spatially localized
patterns that oscillate back and forth across a region in parameter
space (the snaking or pinning region), usually accompanied by repeated
changes in stability. Prototypes for this scenario in one spatial
dimension (1D) are provided by the quadratic-cubic (SH23) and
cubic-quintic (SH35) Swift-Hohenberg equations. These take the form
\huga{\label{sh1} \pa_t u =\lambda u - (1+\pa_x^2)^2u+f(u) } with
$f(u)=a u^2-u^3$ and $f(u)=a u^3-u^5$, respectively, and are
parametrized by the parameters $\lam\in\R$ and $a>0$.  See,
e.g.,~\cite{burke,bukno2007,BKLS09,bbkm2008,dawes08,dawes09,KAC09,hokno2009,SU17}
for basic properties of these equations and their interpretation. In
higher dimensions the situation becomes more complicated, but snaking
behavior of branches of localized patterns can also be
observed. Snaking of localized hexagons in SH23 and related equations
is discussed in, e.g.,~\cite{hexsnake,w18}, while snaking of localized
patterns in 2D reaction-diffusion systems on nonhomogeneous
backgrounds is considered in \cite{schnaki}, and of body-centered cubes 
on homogeneous and nonhomogeneous backgrounds in 3D reaction 
diffusion systems in \cite{w16,UW18b}. Refs.~\cite{kno2008, kno2015} provide reviews of
localization and snaking in various other systems and experiments,
while Refs.~\cite{chapk09,dean11,deWitt19} give analytical results on
homoclinic snaking based on beyond all orders asymptotics.

Motivated by the above considerations we focus here on 1D patterns but consider a more complex nonlinearity $f(u)$ than hitherto studied, in order to study snaking between two distinct periodic patterns. For this purpose we selected the cubic-quintic-septic (SH357) version of \reff{sh1}, namely 
\huga{\label{sh2} 
\pa_t u{=}\lambda u{-}(1{+}\pa_x^2)^2u{+}f(u),\ 
 f(u)={-}au^3{+}bu^5{-}u^7, 
}
and study this equation on bounded domains $\Om=(-l_x,l_x)$ with homogeneous Neumann boundary conditions (BCs) $\pa_x u|_{\pa\Om}= \pa_x^3 u|_{\pa\Om}=0$. The linear stability properties of the trivial solution $u^*\equiv 0$ are independent of $f$ and follow from the linearization $\pa_t v=-(1+\pa_x^2)^2v+\lam v$. This equation has the solutions $v(x,t) = \ee^{ \ii k x+ \mu(k) t }$, $k\in\R$, where $\mu(k) =- (1-k^2)^2+ \lam$. Thus $u^*$ is asymptotically stable for $\lam<0$ and unstable for $\lam>0$ with respect to periodic perturbations with wave number $k$ near $k_c= 1$ and in 1D we expect pitchfork bifurcations to spatially periodic patterns for $\lam\ge 0$, if permitted by the domain and BCs. In detail, if, e.g.,  
$\Om=(-l\pi,l\pi)$, $l\in\N$, then the admissible wave numbers 
are $k\in\frac 1 {2l}\N$, and for large $l$ we have many bifurcation 
points for small $\lam> 0$. The first bifurcation at $\lam_1=0$ has $k_1=1$, and is followed by bifurcations to stripes with $k_{2,3}=1\pm 1/(2l)$, $k_{4,5}=1\pm 1/l, \ldots$, corresponding to sidebands of $k=1$. 

Depending on the parameters $a,b>0$ in \reff{sh2} we find that the bifurcating branches are typically S-shaped, with stable small and stable large amplitude sections, and an unstable intermediate amplitude section.  For definiteness, we choose
\huga{\label{bsel}
b=3.5+0.4(a-3),
}
and consider $a$ as a second free parameter, in addition to $\lam$. We use the following abbreviations:  
\bci
\item $\ups$ for the small amplitude part and $\upl$ for the large amplitude 
part of the periodic branch belonging to wave number $k=1$; 
\item $\upsg$ and $\uplg$ (without specifying a wave number, but typically 
with $k$ near $k_c=1$) for general periodic solutions; 
\item $\uhs$ and $\uhl$ for the spatially homogeneous branch (wave number $k=0$), bifurcating 
at $\lam=1$; 
\item $\um_*$, $*={\rm p}$ or $*={\rm hom}$ for the 
corresponding middle sections. 
\eci 

\begin{figure*}
\begin{tabular}{p{0.48\twi}p{5mm}p{0.48\twi}}
  {\small (a)}
&&{\small (b)}  \\
\ig[width=0.15\twi,height=36mm]{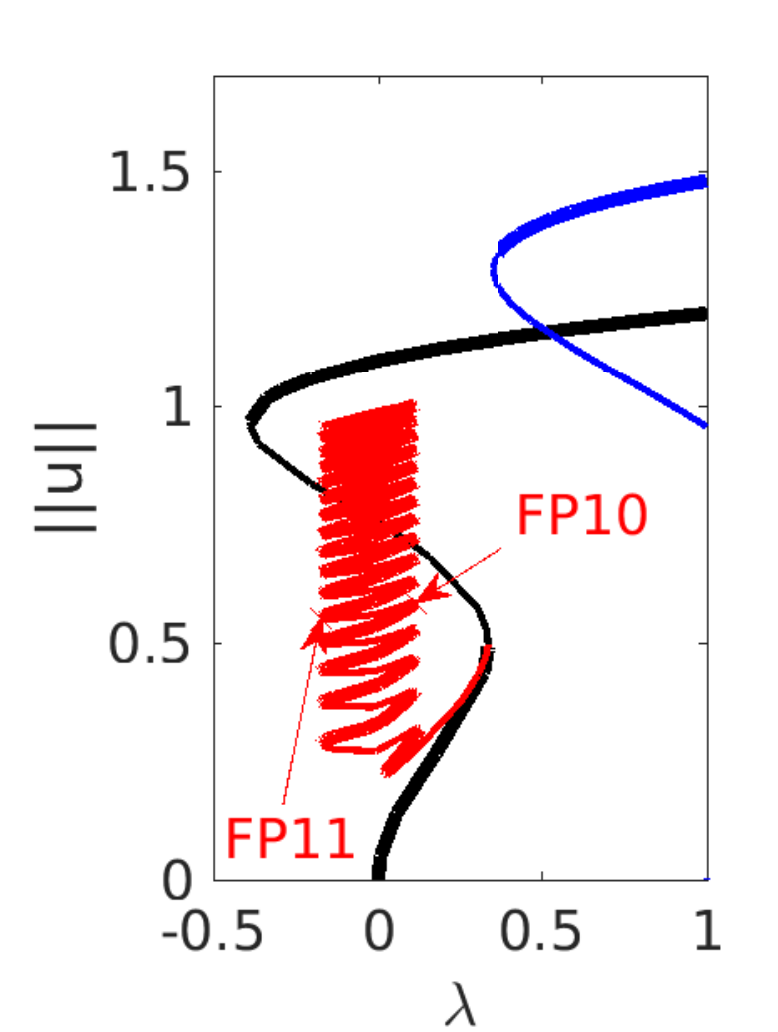}
\hs{-1mm}\raisebox{16mm}{\begin{tabular}{l}
\ig[width=0.17\twi,height=16mm]{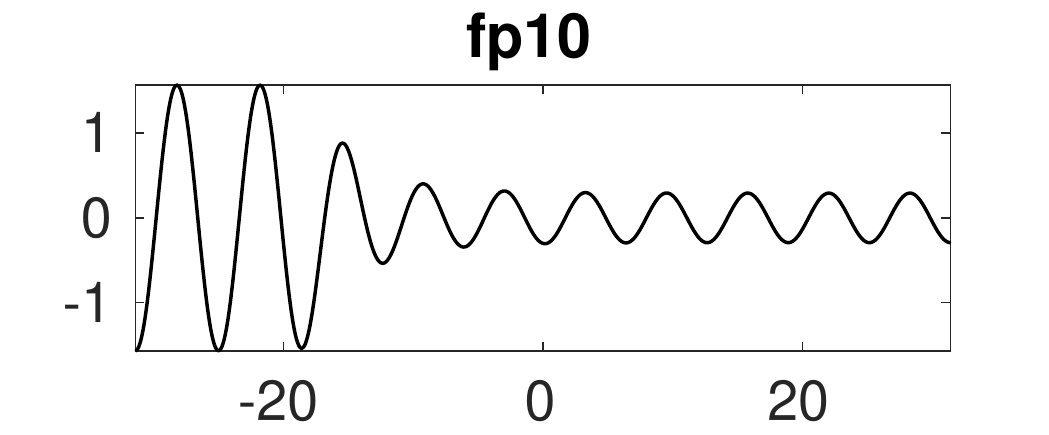}\\
\ig[width=0.17\twi,height=16mm]{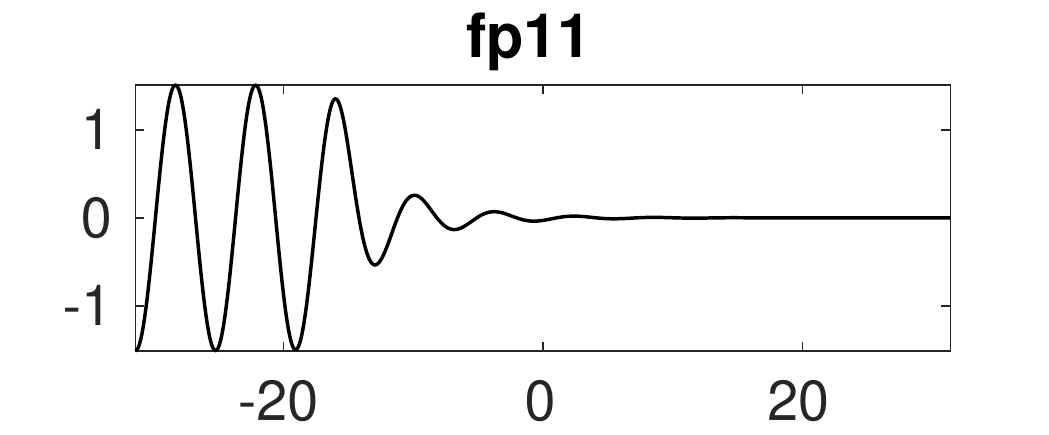}
\end{tabular}}
\ig[width=0.14\twi,height=36mm]{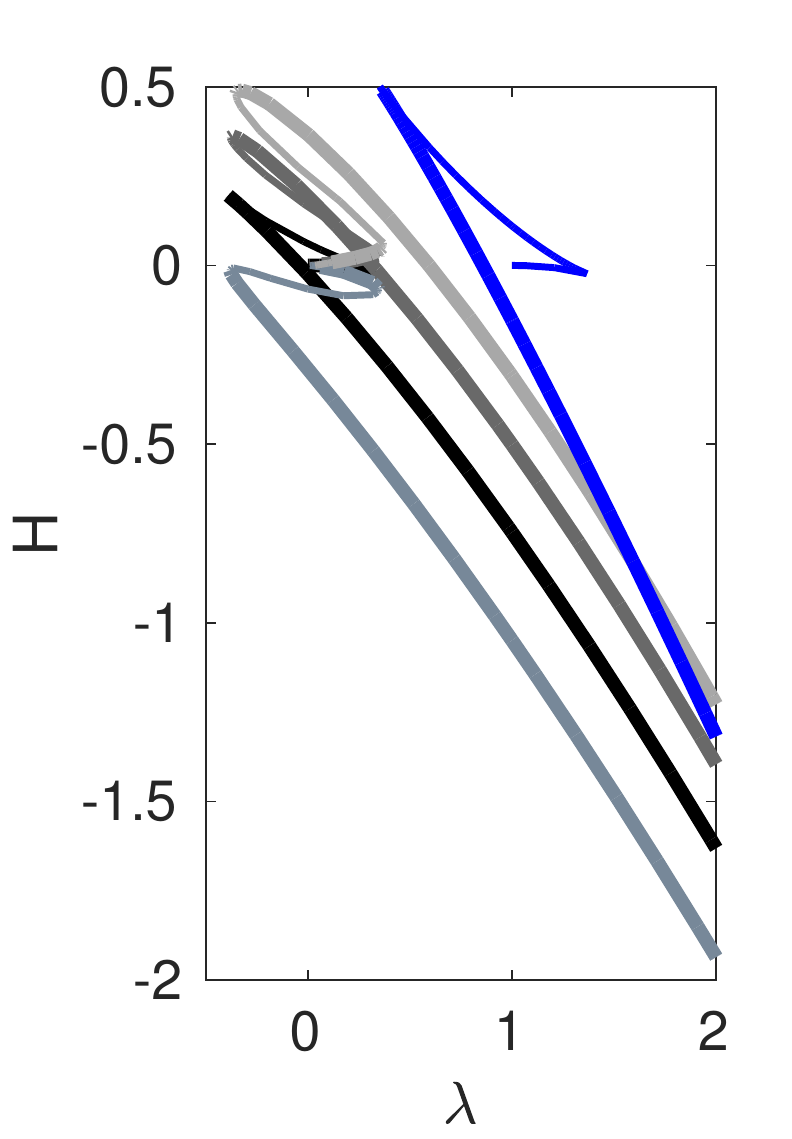}
&&
\ig[width=0.15\twi,height=36mm]{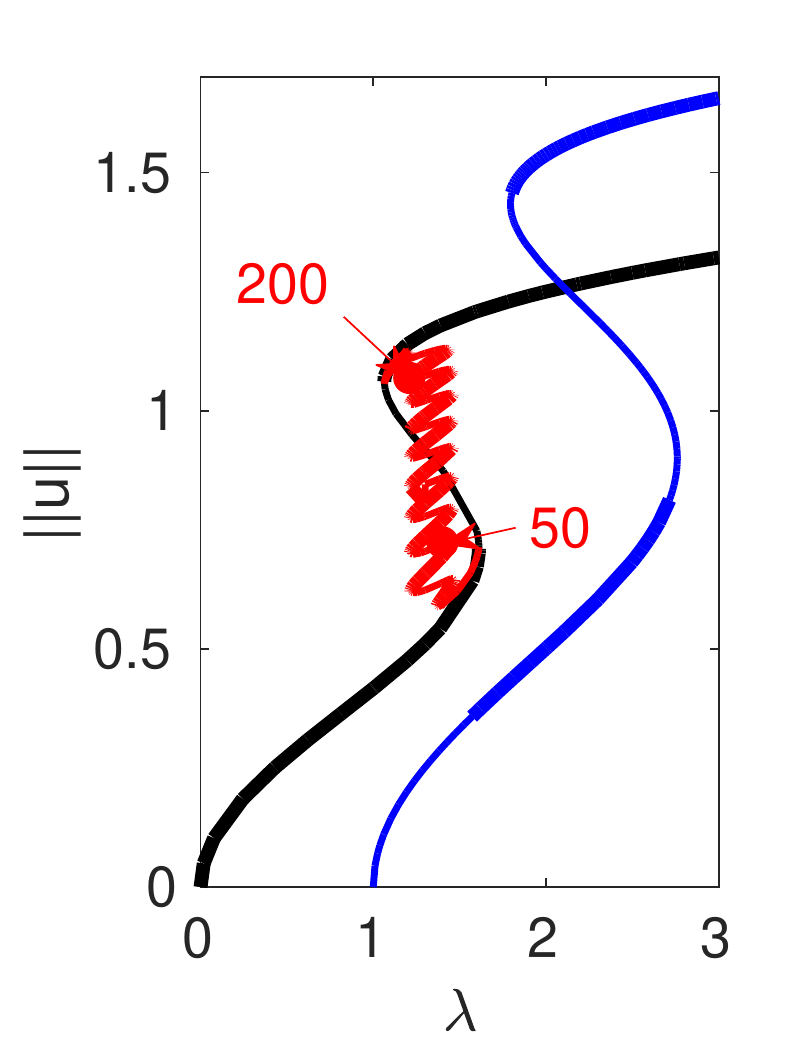}
\hs{-2mm}\raisebox{16mm}{\begin{tabular}{l}
\ig[width=0.17\twi,height=16mm]{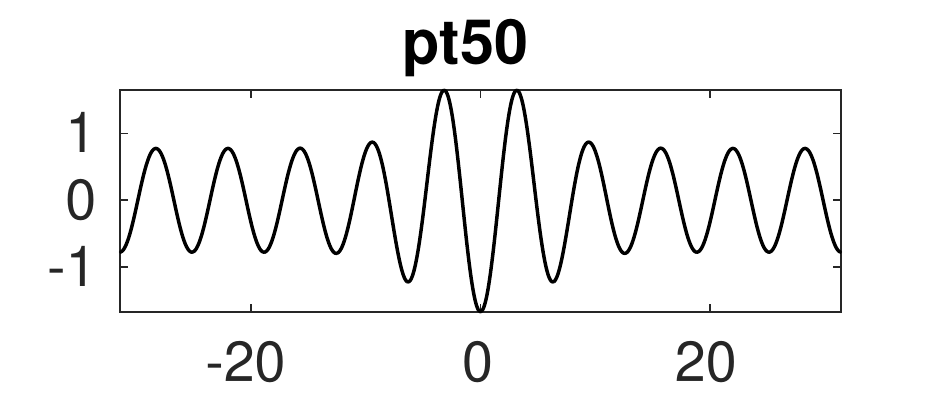}\\
\ig[width=0.17\twi,height=16mm]{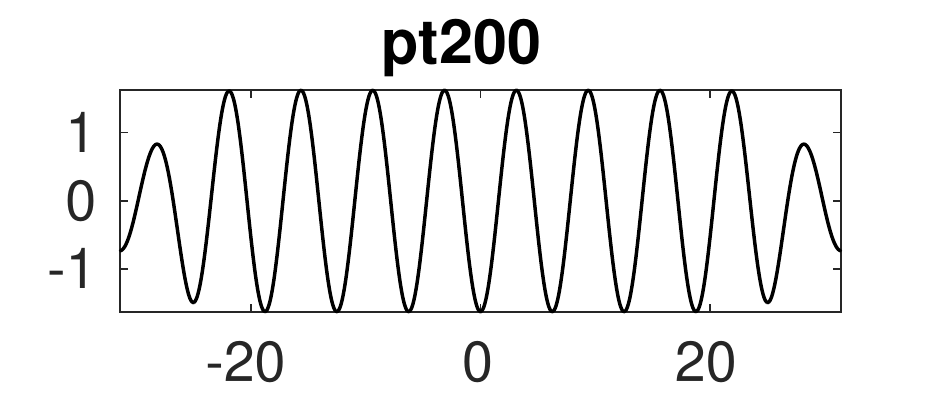}
\end{tabular}}
\ig[width=0.14\twi,height=36mm]{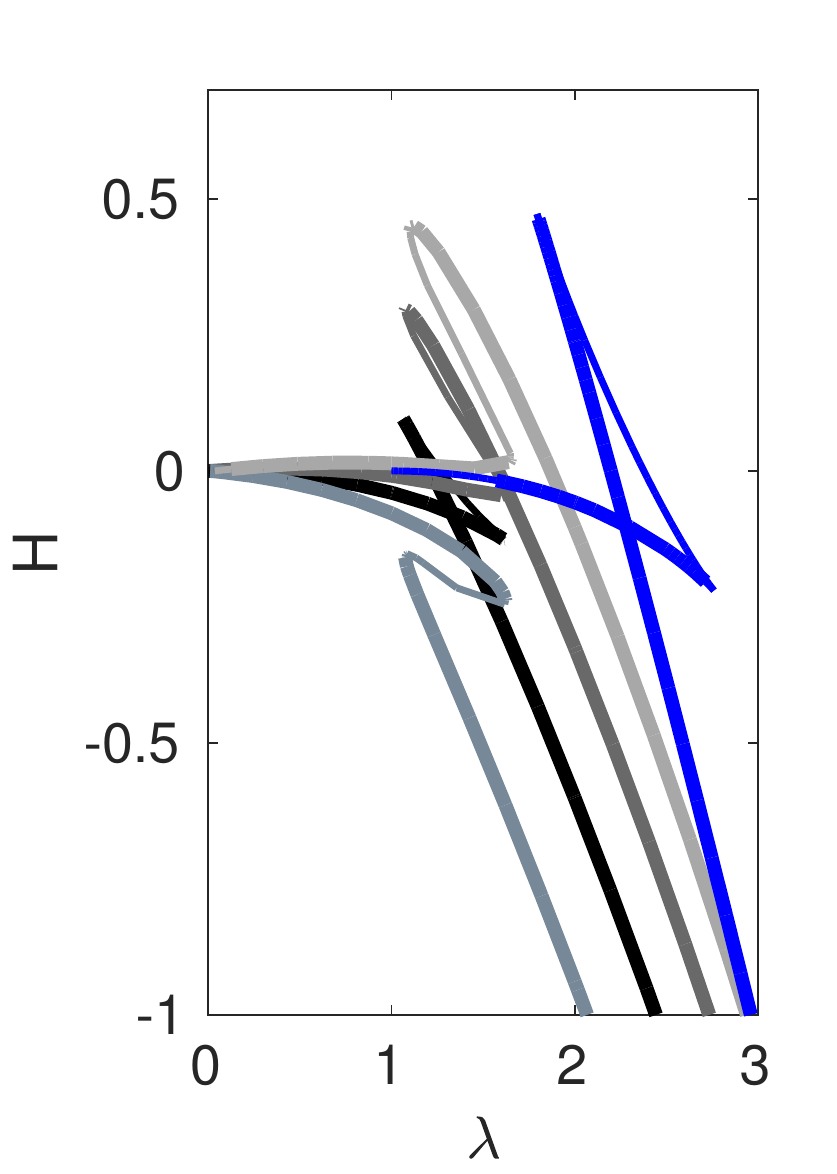}
\\[-1mm]
\end{tabular}
\begin{tabular}{p{0.98\twi}}
  {\small (c)}  \\
\ig[width=0.15\twi,height=45mm]{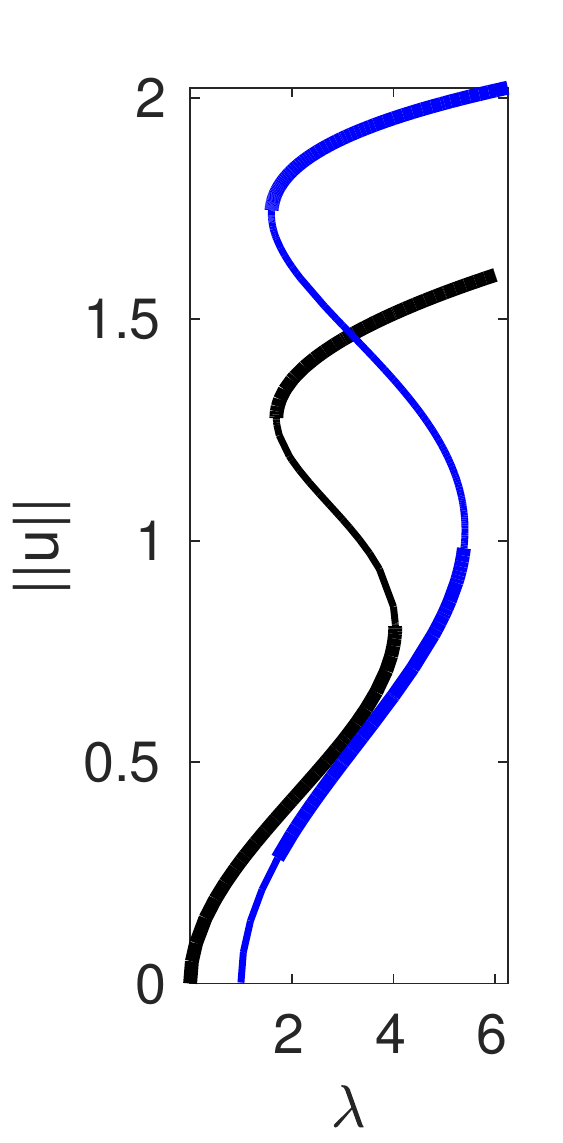}
\ig[width=0.22\twi,height=45mm]{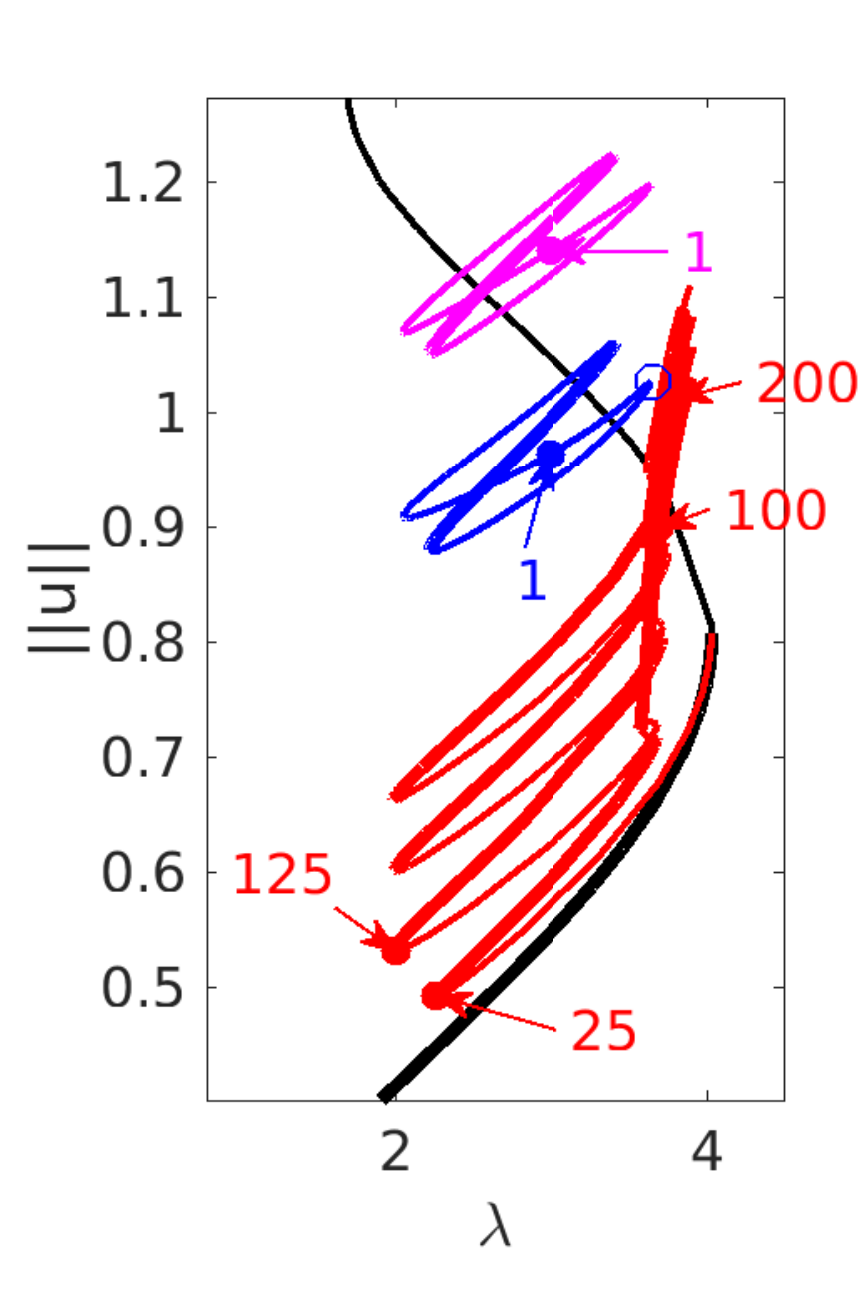}
\hs{-1mm}\raisebox{20mm}{\begin{tabular}{l}
\ig[width=0.2\twi]{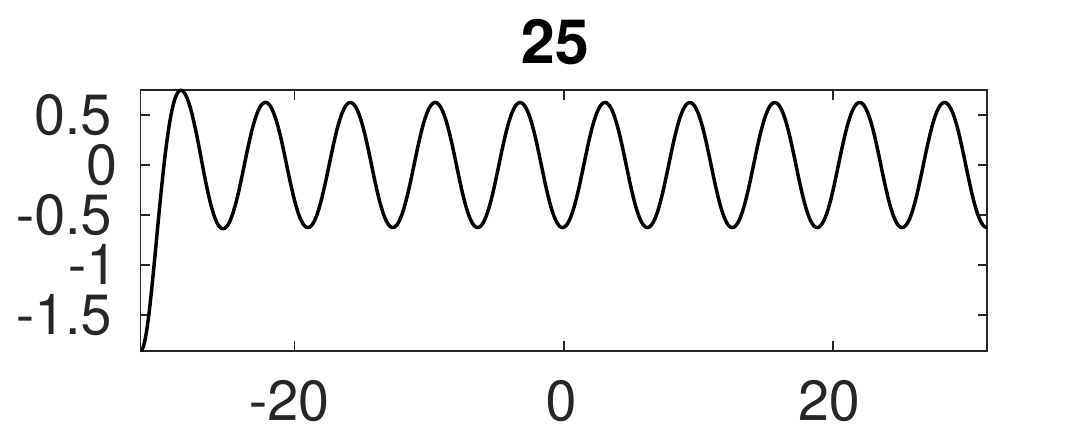}\ig[width=0.2\twi]{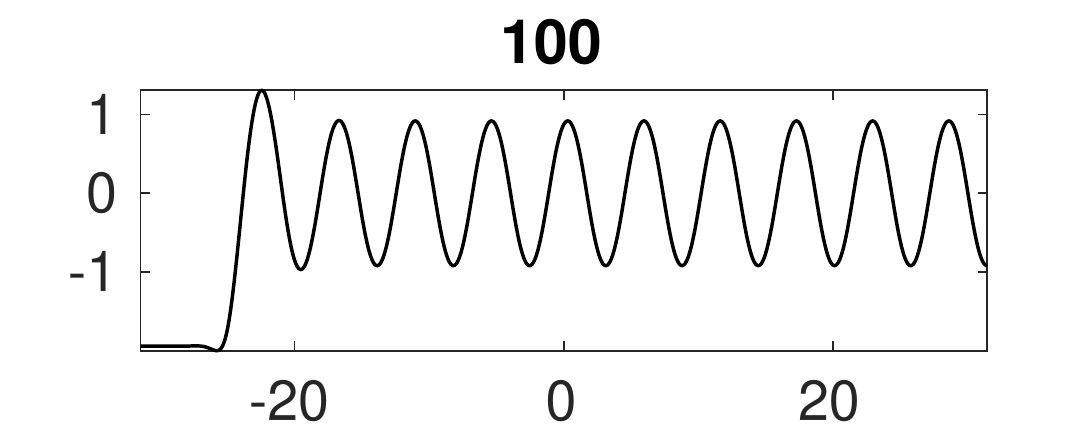}\\
\ig[width=0.2\twi]{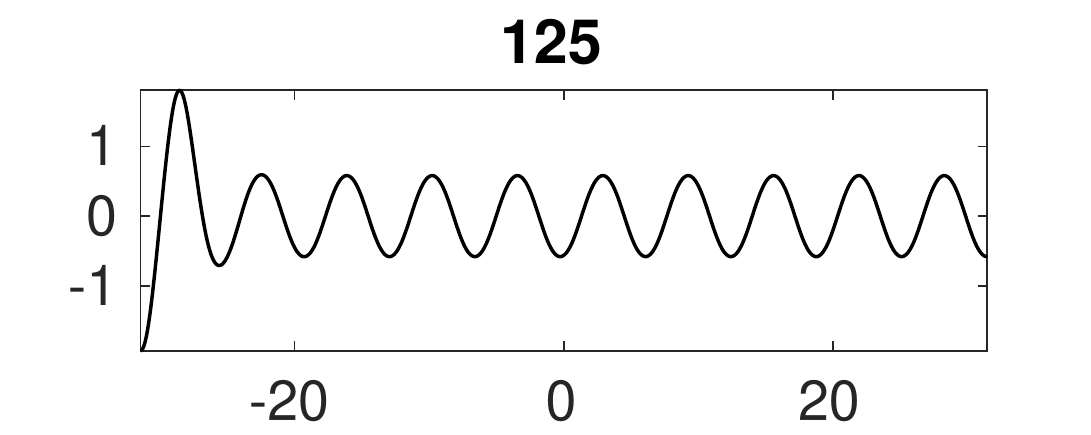}\ig[width=0.2\twi]{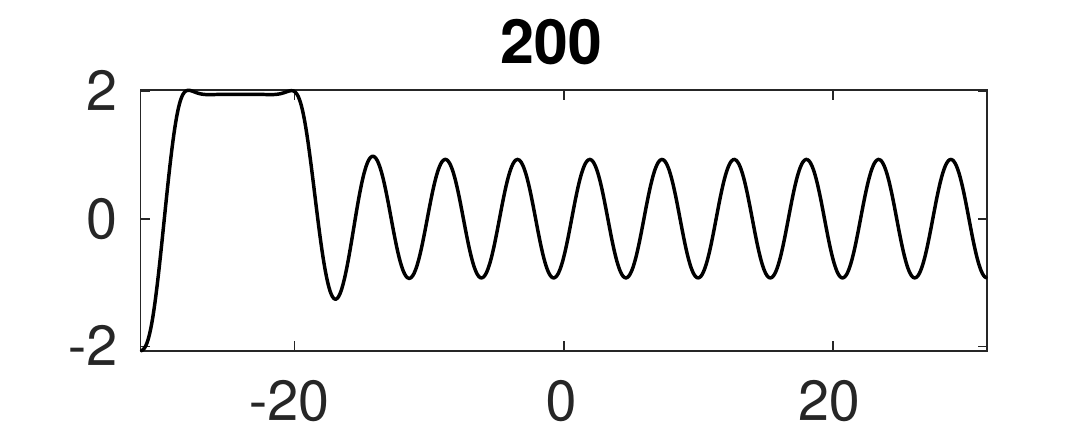}\\
\ig[width=0.2\twi,height=15mm]{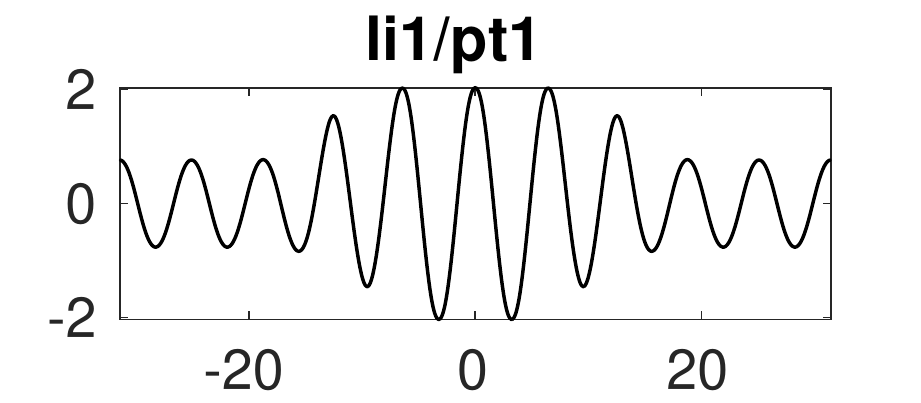}\ig[width=0.2\twi,height=15mm]{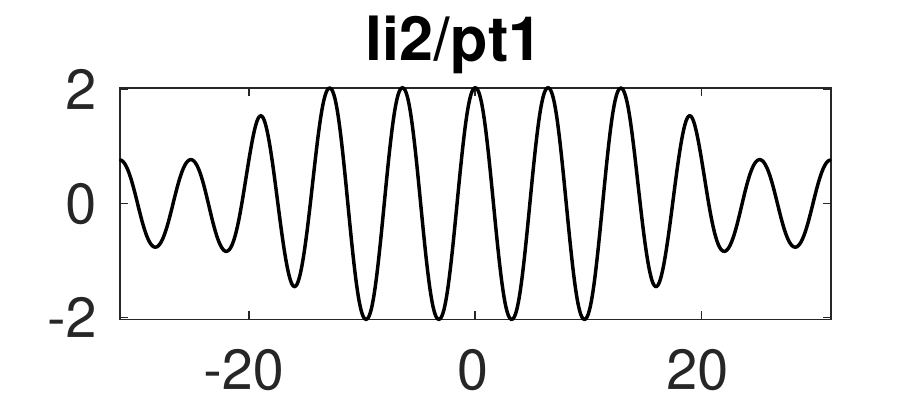}
\end{tabular}}
\hs{-4mm}\ig[width=0.16\twi,height=45mm]{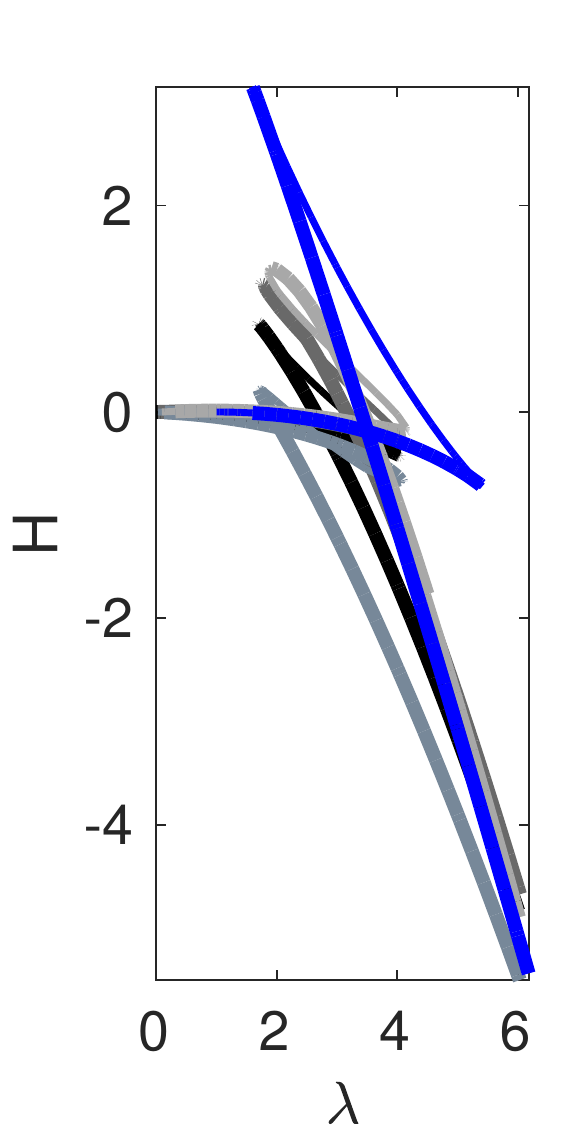}
\end{tabular}
\vs{-2mm}
\caption{{\small Solutions of \reff{sh2} on $\Om=(-10\pi,10\pi)$ with homogeneous Neumann BCs.
(a) $a=2$: bifurcation diagram (BD) of $\up$ (black), $\uhl$ (blue) and one snaking branch (red) 
of fronts between $\upl$ and $u{=}0$ (left folds) and $\uplg$ and $\upsg$ 
(right folds), respectively, showing bistability of $\upl$ with $u=0$ for $\lam<0$, and bistability 
of $\upl$ and $\ups$ for $\lam{>}0$; the snake straddles $\lam{=}0$. 
(b) $a{=}5$: ``classical'' snaking of heteroclinic cycles (red) between $\upsg$ and $\uplg$ (black).
(c) $a{=}9.5$: breakup of classical snaking into stacks of isolas, and a branch of solutions (red) involving $\pm\uhl$ (black). Sample solutions are provided in each case, followed, in the right panels, by plots of the spatial Hamiltonian $H(\cdot)$ showing self-intersections corresponding to different types of Maxwell points. Thick (thin) lines in the 
bifurcation diagrams and plots of $H$ indicate linearly stable (unstable) solutions; 
see also Remark \ref{rem1}(d) for the naming and plotting conventions in these and the following plots.
}}\label{f0}
\end{figure*}

Figure \ref{f0} provides an overview of the dependence of the solution sets of \reff{sh2} on $a$, and introduces the solution types considered in this paper. Thick lines indicate linearly stable states while thin lines correspond to linearly unstable states. For $a>0$ and not too large the S-shape of the first bifurcating branch is mild, and the subsequent branches are 'neatly ordered' in the sense that only a short interval of the small amplitude periodic solutions is stable, and there is no overlap of the bistable range of the periodic solutions with wave number $k$ near 1 with the branch of spatially homogeneous solutions (wave number $k=0$) [Fig.~\ref{f0}(a) for $a=2$]. If we increase $a$, the S-shape and hence the overlap of the bistable ranges for solutions with different $k$ become more pronounced [Fig.~\ref{f0}(b) for $a=5$]. This overlap is of interest to us because we expect localized patterns to be generically present within such bistable (or indeed multistable) ra
 nges.

It turns out that for the smaller $a$ one finds 'almost classical' snaking of localized states
that is associated with \hecy s between $\upsg$ with wave numbers close to $k=1$, and
$\uplg$, again with wave numbers close to $k=1$. These states consist of a portion of
$\uplg$ in a background of $\upsg$ or vice versa; localized states consisting of a
portion of $\uplg$ in $u=0$ background are also possible. For larger 
$a$, more and more branches (with wave numbers deviating from $k=1$) 
enter the game, including the branch $\uh$ of spatially homogeneous 
solutions ($k=0$), and the solution set of \reff{sh2} becomes more and 
more complicated. For instance, the snaking branches of small--to--large 
periodic patterns break up into isolas, and additional branches consisting 
of \hecy s between various distinct spatial patterns enter the picture. 
Figure \ref{f0} is thus intended as a preview of the subsequent results.
The norm $\|u\|$ used in (a) and all similar plots is 
\huga{ \|u\|:=\left(\frac 1 {|\Om|}\int_\Om u^2\dd x\right)^{1/2} 
\text{ (normalized $L^2$ norm).}
}

In summary, in this paper we numerically investigate how the set 
of localized patterns of \reff{sh2} becomes richer and richer 
with increasing $a$, and literally 'explodes' for $a\approx 9.5$ and larger. 
These results are presented in detail in \S\ref{rsec}, while \S\ref{dsec} 
provides a brief discussion. 

\brem\label{rem1}
(a) Equation \reff{sh2} has a number of symmetries: (i) translational
invariance (for $\Om=\R$); (ii) odd symmetry $u\mapsto -u$; (iii) spatial reflection
symmetry $x\mapsto-x$. The translational invariance (i) is 
broken over $\Om=(-l_x,l_x)$ by the Neumann BCs $\pa_x u|_{\pa\Om}=\pa_x^3 u|_{\pa\Om}=0$, but ``periodic''
solutions over $(-l_x,l_x)$ can be extended to all of $\R$ by reflection at the boundaries. Thus 
the first 'front' in Fig.~\ref{f0}(a) can also be seen as heteroclinic cycle
between a large amplitude and a small amplitude periodic solution. 
(ii) implies that all nontrivial solutions are double, and we generally identify 
$\pm\upsg$, $\pm\uplg$, and $\pm \uh$, respectively. 
(ii) and (iii) together imply
that we have branches of {\em odd solutions} of the form $u(-x)=-u(x)$, as 
opposed to {\em even solutions} which have a maximum or minimum at $x=0$. 
As a consequence, snaking branches come in odd and even families, and 
generally we expect ladder branches connecting these, and this {\em snakes--and--ladders structure} is a prerequisite for the breakup of snakes into isolas.
These results are common to SH35 \cite{burkeK}.

(b) Equation \reff{sh2} is a gradient system, $\pa_t u=-\nabla \CE(u)$, with respect to
the energy 
\huga{\label{en1}
\CE(u)=\int_\Om \frac 1 2 ((1+\Delta)u)^2-\frac 1 2 \lam u^2-F(u)\dd x, 
} 
$F(u)=\int_0^uf(v)\dd v$, 
where either $\Om=\R$, or $\Om=(-l_x,l_x)$ with 
Neumann BCs $\pa_x u|_{\pa\Om}=\pa_x^3 u|_{\pa\Om}=0$. 
In particular, local minima of $\CE$ are stable stationary solutions of \reff{sh2}, and \reff{sh2} does not have time--periodic solutions (with finite energy). 
Moreover, the translational invariance of $\CE$ yields the 
existence of a spatially conserved quantity for steady solutions, 
a spatial Hamiltonian, cf., e.g., \mbox{\cite[Proposition 1]{strsnake},} here given 
by   
\huga{
H(u)=\pa_xu\pa_x^3u-\frac 1 2 (\pa_x^2u)^2+(\pa_x u)^2+\frac 1 2 (1{-}\lam)u^2-F(u). \label{ham1d}
}
Hence, a necessary condition for a heteroclinic connection between, e.g., two periodic
solutions $\upsg$ and $\uplg$ is that $H(\upsg)=H(\uplg)$. 
For the classical SH23 and SH35 equations, this requirement provides an important 
wave number selection principle that determines the wave number $k(\lam)$ along the snaking
branches. The same is true for SH357. Figure \ref{f0}(e) 
indicates that while for small to moderate values of $a$ there are few intersections of $H$ 
for the different branches, this is no longer so for larger $a$ where the number of possible heteroclinic cycles becomes very large.

(c) When choosing the bounded domain $\Om$ for Equation \reff{sh2} we need to compromise between 
\bcen
\item Generality: the results should be representative of the situation on 
large domains, ideally approximating the case $\Om=\R$; this in general calls 
for large domains. 
\item Feasibility and clarity: the domain should be small enough to 
(i) avoid exceedingly expensive numerics, and (ii) allow clear plotting 
of results. 
\ecen 
It turns out that the basic results can be studied and presented well on relatively small domains, for instance $\Om=(-10\pi,10\pi)$, which we use in most cases. In some cases, in particular for extracting wave numbers from 
mixed periodic solutions via Fourier transform, a larger $\Om$ such 
as $\Om=(-30\pi,30\pi)$ is helpful. In any case, we checked that none 
of our results depends qualitatively on the domain size by running 
the same numerics on significantly larger domains, where however the 
results become more difficult to present graphically. 

(d) For the plots of bifurcation diagrams (BDs) and sample solutions 
we use the following conventions. Stable parts of branches (as determined 
from the eigenvalues of the linearization around solutions) are plotted 
as thick lines, unstable parts as thinner lines. Dots labeled by an integer $n$
correspond to solutions for which we plot profiles $u(x)$, titled 
``{\tt pt}$n$'' if there is no ambiguity. Fold and branch points are indicated via {\tt FP}$n$
and {\tt BP}$n$, respectively, 
and similarly in the title of the sample plots. Occasionally we give titles 
in the form {\tt branch/point}. 

(e) When $\Omega=\R$ the spatial dynamics picture implies that $u=0$ is a saddle for $\lam<0$ with
two stable and two unstable eigenvalues. Likewise, a robust connection to $\upsg$ or $\uplg$ require these to be generalized saddles and hence that they have a three-dimensional center-stable manifold and a three-dimensional center-unstable manifold. This is a consequence of spatial reversibility and the conservation of $H$. 
  \eex 
\erem


\section{Results}\label{rsec}

We use \pdep\ \cite{p2p, p2phome} to compute bifurcation diagrams for \reff{sh2}.  As domain we typically choose $\Om=(-10\pi,10\pi)$, which is large enough to permit a multitude of patterns, cf.~Remark \ref{rem1}(c). 

\subsection{The case $a=2$}\label{ss2}

We start with $a=2$. In the plot of $H$ as a function of $\lambda$ in Fig.~\ref{f0}(a) we see that there 
are few self-intersections for the first four bifurcating branches, 
and in particular no overlap of their bistable ranges with the 
spatially homogeneous (blue) branch. This corresponds to the 'easy' 
situation where relatively few heteroclinic connections are possible. 
Moreover, in this case transitions between different heteroclinic cycles 
and interesting codimension two points can be easily identified. 

\begin{figure}[ht]
\bce 
\begin{tabular}{ll}
(a)&(b)\\ 
\ig[width=0.18\twi,height=45mm]{2-L2}&
\hs{-0mm}\raisebox{22mm}{\begin{tabular}{l}
\ig[width=0.23\twi,height=16mm]{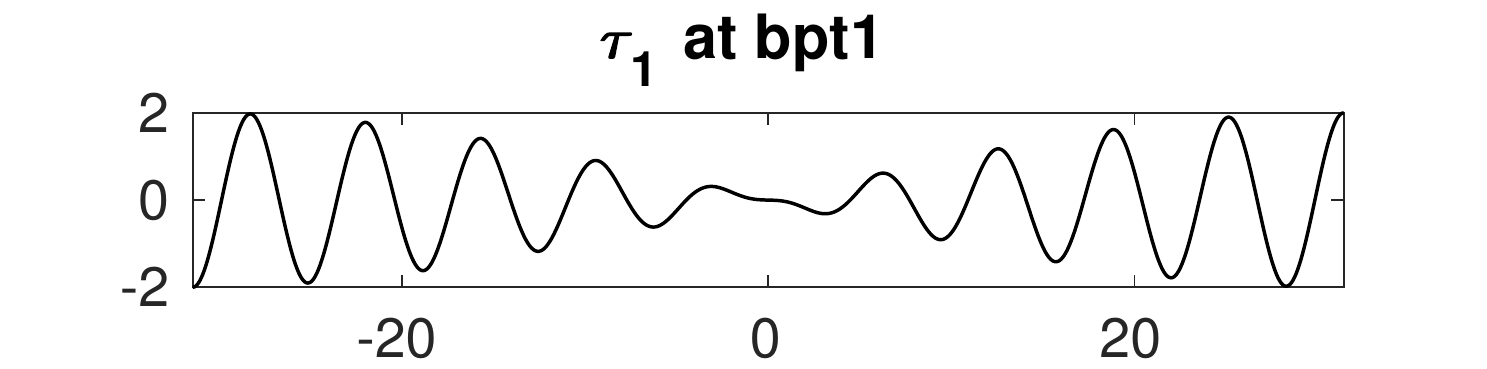}\\
\ig[width=0.23\twi,height=14mm]{2-fp10}\\
\ig[width=0.23\twi,height=14mm]{2-fp11}
\end{tabular}}\\
(c)&(d)\\
\hs{-0mm}\ig[width=0.18\twi,height=50mm]{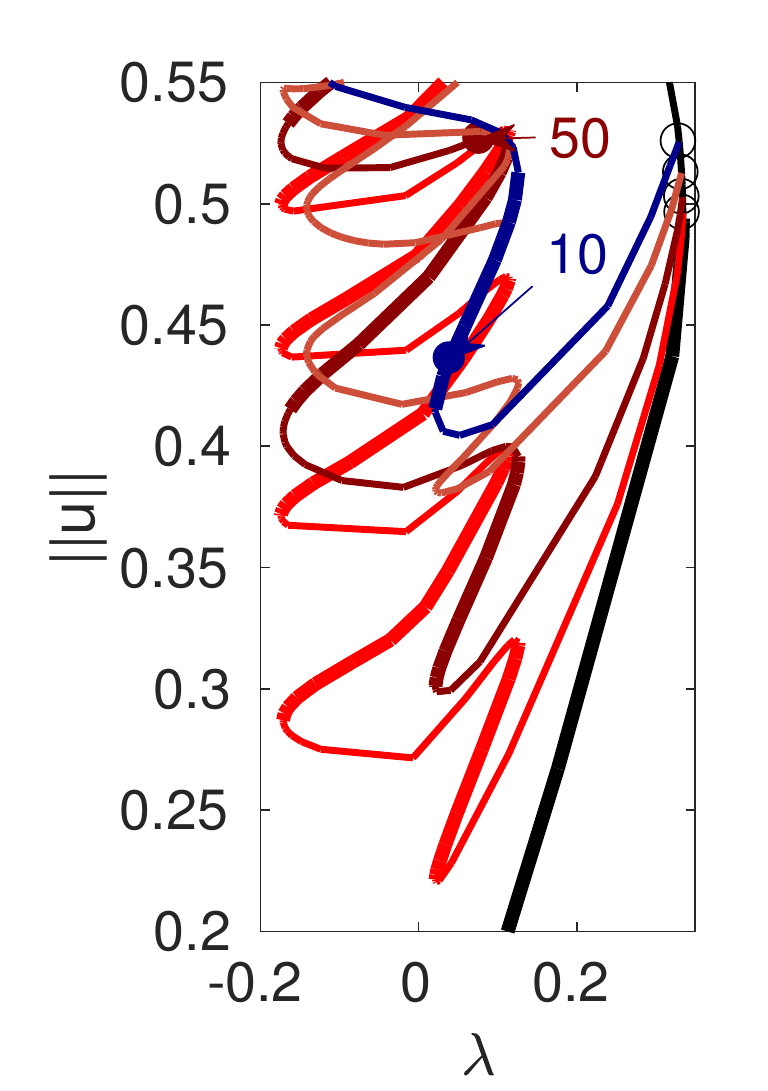}&
\hs{-0mm}\raisebox{25mm}{\begin{tabular}{l}
\ig[width=0.23\twi,height=13mm]{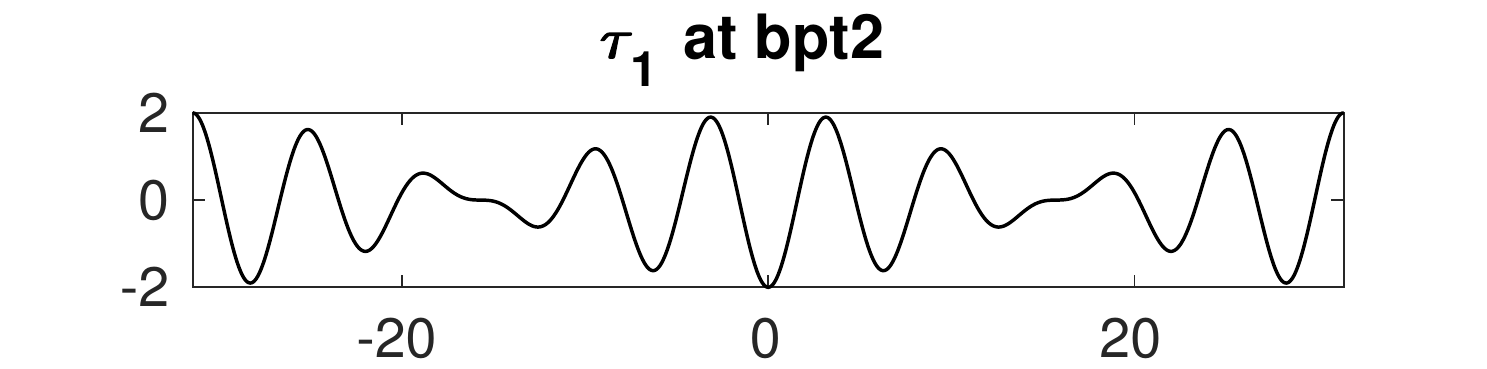}\\
\ig[width=0.23\twi,height=13mm]{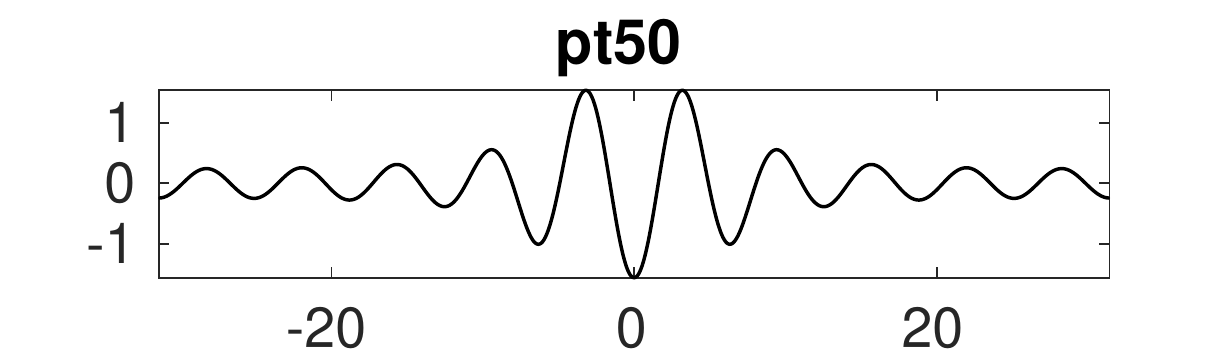}\\
\ig[width=0.23\twi,height=13mm]{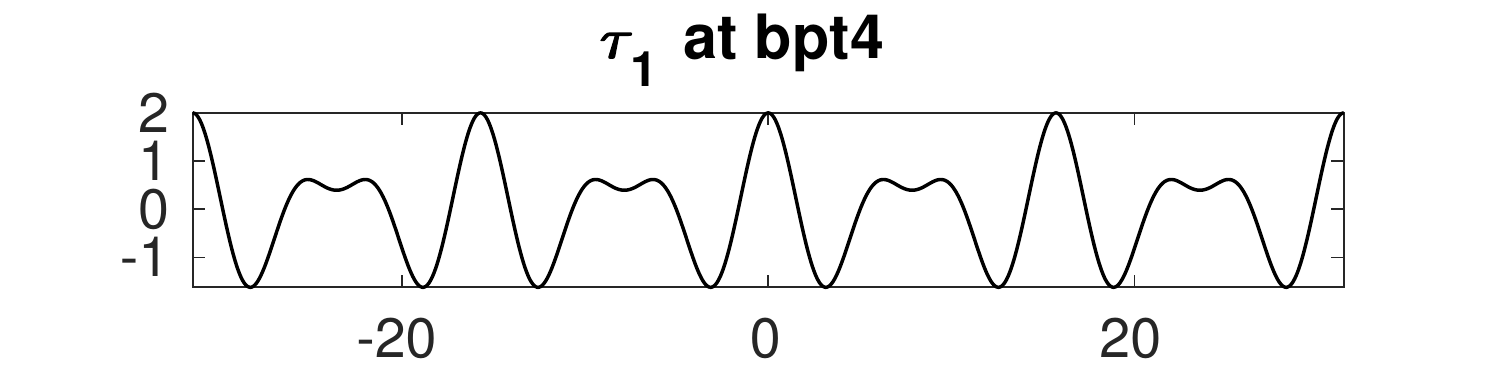}\\
\ig[width=0.23\twi,height=13mm]{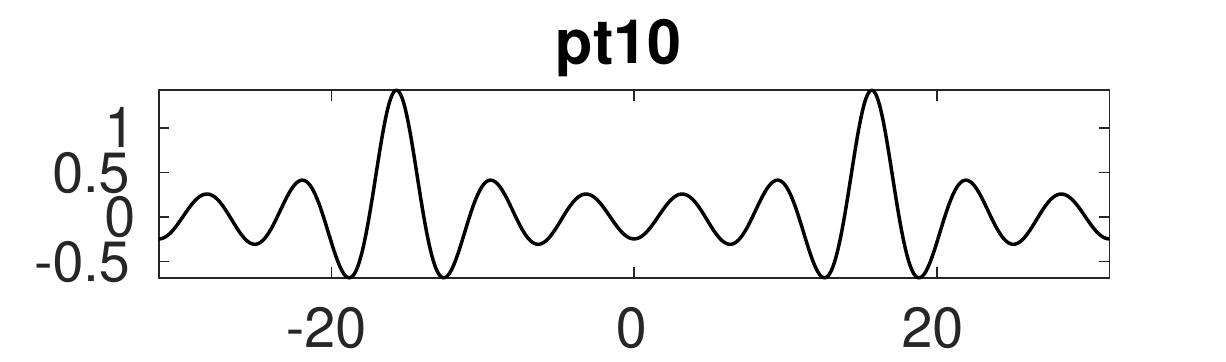}
\end{tabular}}
\end{tabular}
\ece
\vs{-2mm}
   \caption{{\small $a=2$. (a,b) Snaking fronts between $\uplg$ and 
$\upsg$ (for $\lam>0$) resp.~between $\uplg$ 
and $u=0$ (for $\lam\le 0$, where $\upsg$ does not exist). (c) Zoom into 
the BD near the first fold on $\up$. After the fold there are multiple 
BPs. The front solutions bifurcate at BP1 [red branch in (a)]. 
The first plot in (b) shows the tangent direction for this bifurcation. 
Branches of solutions with multiple interfaces bifurcate at subsequent BPs.
For instance, a $\upsg-\uplg-\upsg$ \hecy\ bifurcates at BP2, 
and a \hecy\ with two 'pulses' at BP4; see (d) for sample plots.  \label{f2}}}
\end{figure}

Figure \ref{f2} provides more details. In particular, we find that 
after the first fold on $\up$ there are multiple BPs where branches 
of \hecy s with an increasing number of interfaces bifurcate. The first 
branch corresponds to fronts (but see also Remark \ref{rem1}(a) for the 
interpretation of the front as a \hecy\ via 
even parity continuation of the solution over the domain boundary), the second to a $\upsg-\uplg-\upsg$ \hecy, 
and so on. In the following we concentrate on the second (pulse-like) branch for moderate values of $a$ but a particular feature of the small $a$ regime is clearly visible on the first branch: for $\lam<0$, $\uplg$ connects to the trivial state $u=0$, while for $\lam>0$ it connects to $\upsg$. In other words, each time $\lambda$ passes through $\lambda=0$ into $\lambda>0$ the hole in the solution fills in with small amplitude oscillations. On an infinite domain the \hecy\ changes from one involving the $u=0$ state to one involving $\upsg$. Note that there will be parameter values such that the right folds of the snaking front branch just reach $\lambda=0$, a situation corresponding to a codimension two bifurcation of \hecy s.
There is a second codimension two transition that is relevant as well: when $\upsg$ changes from subcritical to supercritical the termination point of the localized solutions moves from $\lam=0$ to the right fold on $\upsg$ as in Fig.~\ref{f2}(a). An analogous transition has been observed in rotating plane Couette flow 
\cite{SGS19}. 

The front solutions bifurcate from BP1, the branching point nearest to the right fold of $\upsg$; other branching points, further away from the fold, give rise to solutions with multiple interfaces, as summarized in Fig.~\ref{f2}(c,d).

\subsection{The case $a=5$}\label{ss5}

\subsubsection{Snakes and ladders of cycles between $\ups$ and $\upl$}
For $a=5$, the second (leftmost) fold on $\up$ is at $\lam>0$, and the snakes bifurcating near the first fold lie entirely in the $\lam>0$ range. The red branch in Fig.~\ref{f1}(b) shows the (even) $\upsg-\uplg-\upsg$ snake $L_0$ bifurcating at BP2. Looking more closely, one finds secondary bifurcation points near every fold on this branch giving rise to 'rungs' [brown branches in (c)] connecting to the odd snake [$L_{\pi/2}$, blue branch in (c)]. 
This ladder structure becomes important for understanding the breakup 
of snakes at larger $a$, see \S\ref{ss95}. 
Additionally, in Fig.~\ref{f1}(b) we plot a branch connecting two Turing bifurcation
points on $\uh$. Note that except for $\ups$, all nontrivial branches are unstable at 
bifurcation but many become stable at (small but) finite amplitude.

\begin{figure}[H]
\bce 
\begin{tabular}{ll}
(a)&(b)\\
\ig[width=0.2\twi,height=40mm]{5-H}&
\ig[width=0.17\twi,height=40mm]{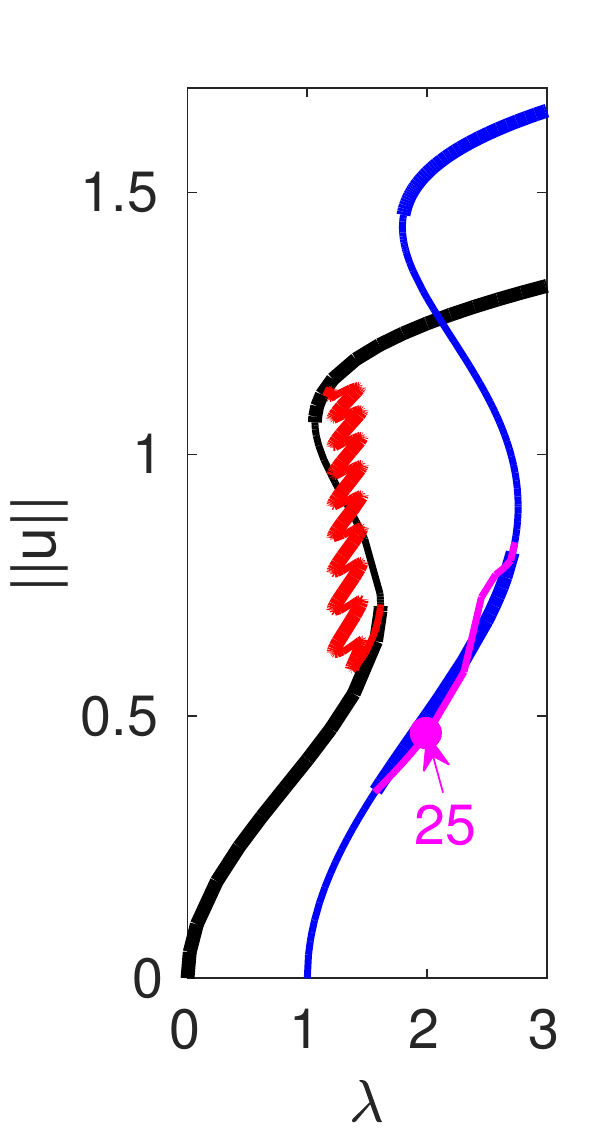}\\
(c)&(d)\\
\hs{-0mm}\ig[width=0.22\twi,height=45mm]{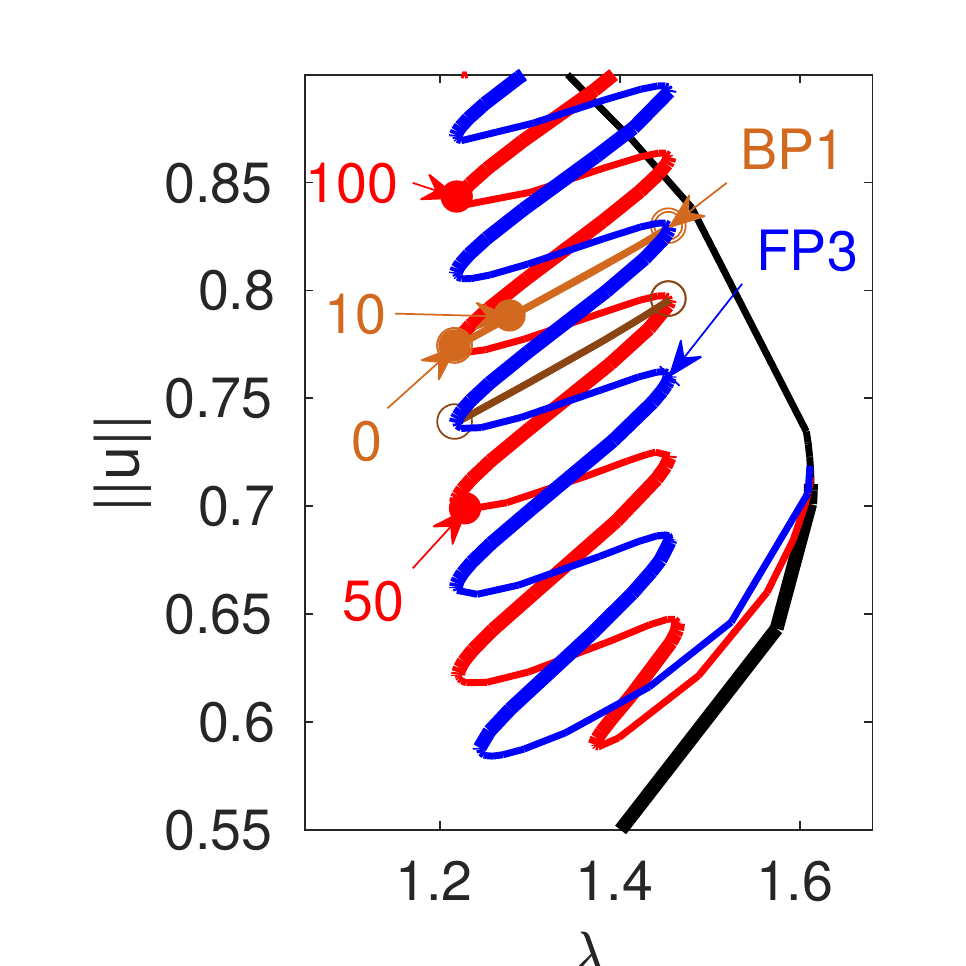}&
\hs{-2mm}\raisebox{22mm}{
\begin{tabular}{l}
\ig[width=0.22\twi,height=13mm]{5-50}\\
\ig[width=0.22\twi,height=13mm]{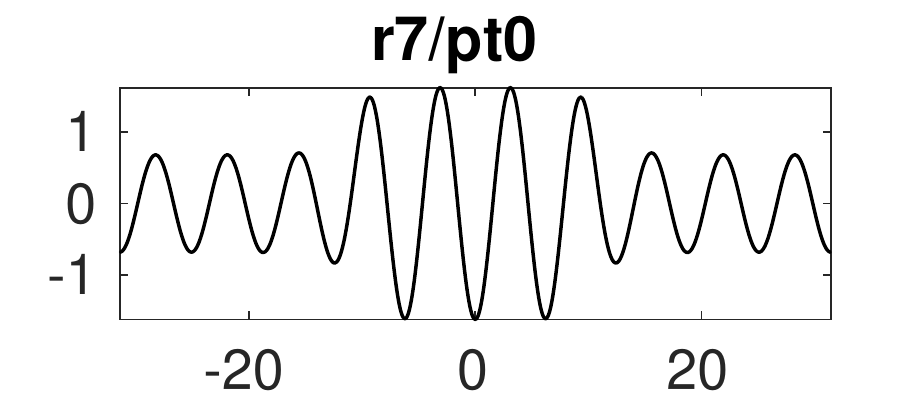}\\
\ig[width=0.22\twi,height=13mm]{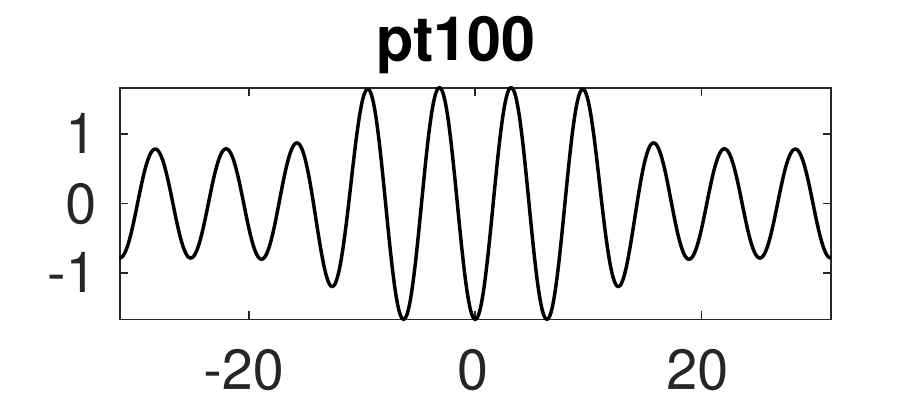}
\end{tabular}}
\end{tabular}\\[-0mm]
\begin{tabular}{l}
(e)\\
\ig[width=0.22\twi,height=13mm]{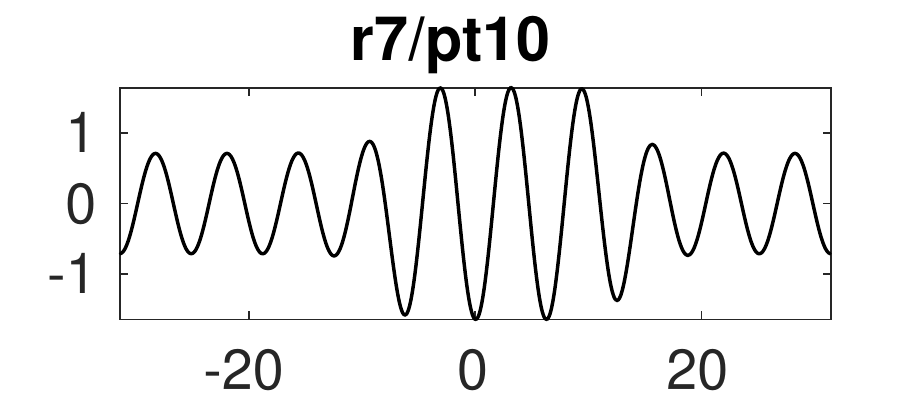}
\ig[width=0.22\twi,height=13mm]{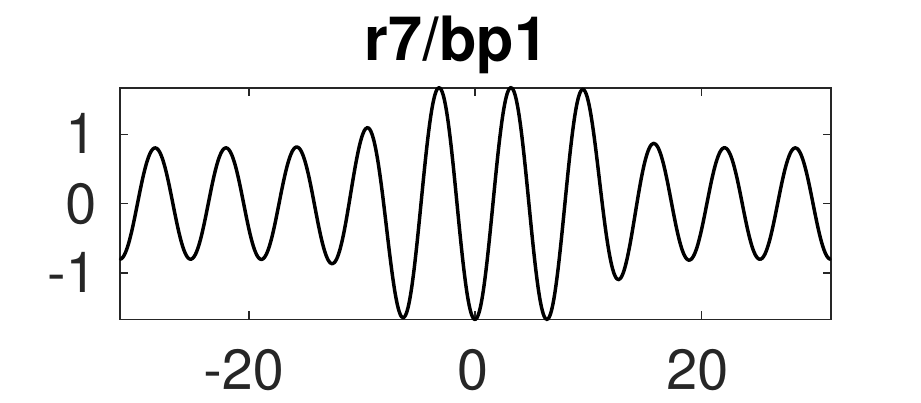}\\
\hs{-0mm}\ig[width=0.22\twi,height=13mm]{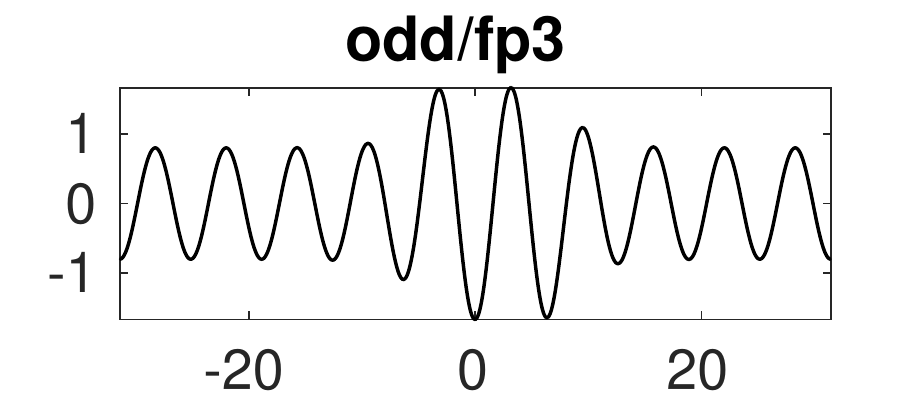}
\ig[width=0.22\twi,height=13mm]{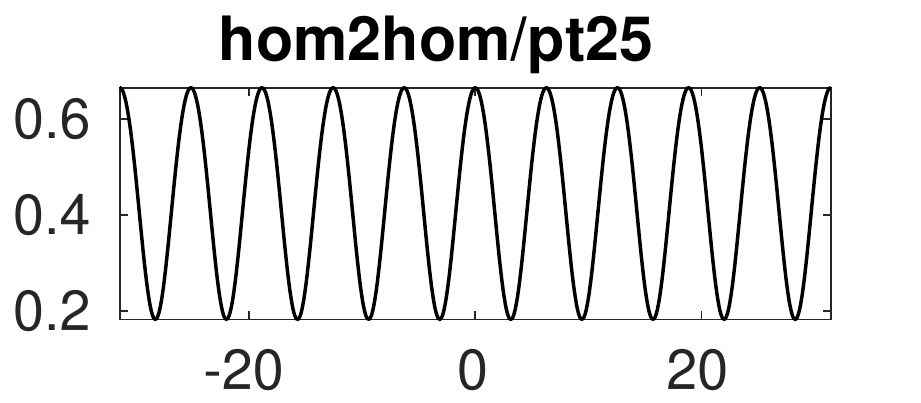}
\end{tabular}\\[-2mm]
\ece
\vs{-2mm}
   \caption{{\small 
$a=5$. (a) $H$ on the first four branches of periodic solutions (black to light grey), 
and the homogeneous solution $\uh$ ($k=0$, blue). (b) $\|u\|$ for the
primary periodic branch (black) and for $\uh$ (blue), together with a snake (red) of cycles between $\upsg$ and $\uplg$, consisting of even solutions, and a pattern branch between two Turing bifurcations on $\uhl$ (magenta). (c) Zoom of the snake, together with two rungs (brown) that connect the even snake and the odd snake (blue). (d) Even solutions (close to left folds) corresponding to (c).  (e) Sample solutions from the light brown rung and the odd parity snake (blue branch) in (c), and from the magenta branch ({\tt hom2hom}) in (b). 
  \label{f1}}}
\end{figure}

\subsubsection{Continuation in the domain size}
Along the snaking branches of, e.g., $\upsg-\uplg-\upsg$ \hecy s, the wavelengths (and hence amplitudes) of both
$\upsg$ and $\uplg$ must in general change continuously as determined by the condition
$H(\upsg(\cdot;\lam))\stackrel!=H(\uplg(\cdot;\lam))$. Here we look into this phenomenon
in more detail, via continuation in the domain size scaling $\ell$. Thus we 
modify \reff{sh1} to 
\huga{\label{sh1b} 
\pa_t u =\lambda u - (1+\ell^{-2}\pa_x^2)^2u+f(u), 
} 
on $\Om=(-l_x,l_x)$, such that the effective domain is $\Om_\ell:=\ell\Om=(-\ell l_x,\ell l_x)$. 
Qualitative results are: 
\bci 
\item[(a)] In general, $\uplg$ is more rigid than $\upsg$. This means 
that $\uplg$ adapts its wave number and amplitude less than $\upsg$ does. \item[(b)] 
The continuation in $l$ leads to a new kind of snaking, where the small 
amplitude part of the \hecy s grows or shrinks via phase slips in $\upsg$.  
\eci 
These results are illustrated in Figs.~\ref{fl1} and \ref{fl2}. Here, 
to accurately extract the wave numbers of the periodic patterns we choose 
a rather large base domain $\Om=(-30\pi,30\pi)$, although the phenomenon 
can be observed on significantly smaller domains, e.g., $\Om=(-10\pi,10\pi)$. 

\begin{figure*}[htpb]
\bce 
\begin{tabular}{lll}
(a)&(b)&(c)\\
\hs{-0mm}\ig[width=0.21\twi]{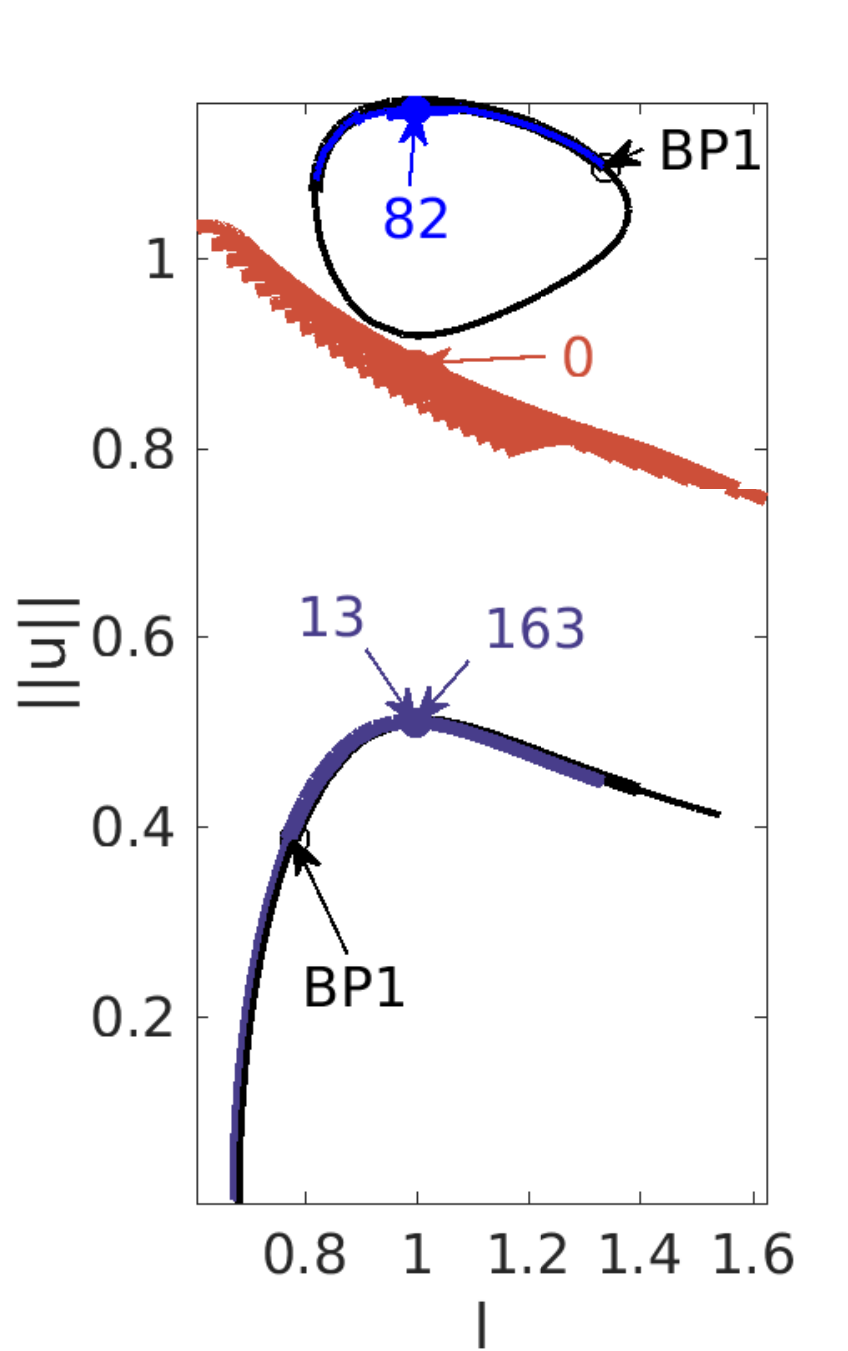}&
\hs{-0mm}\raisebox{30mm}{
\begin{tabular}{l}
\ig[width=0.4\twi,height=17mm]{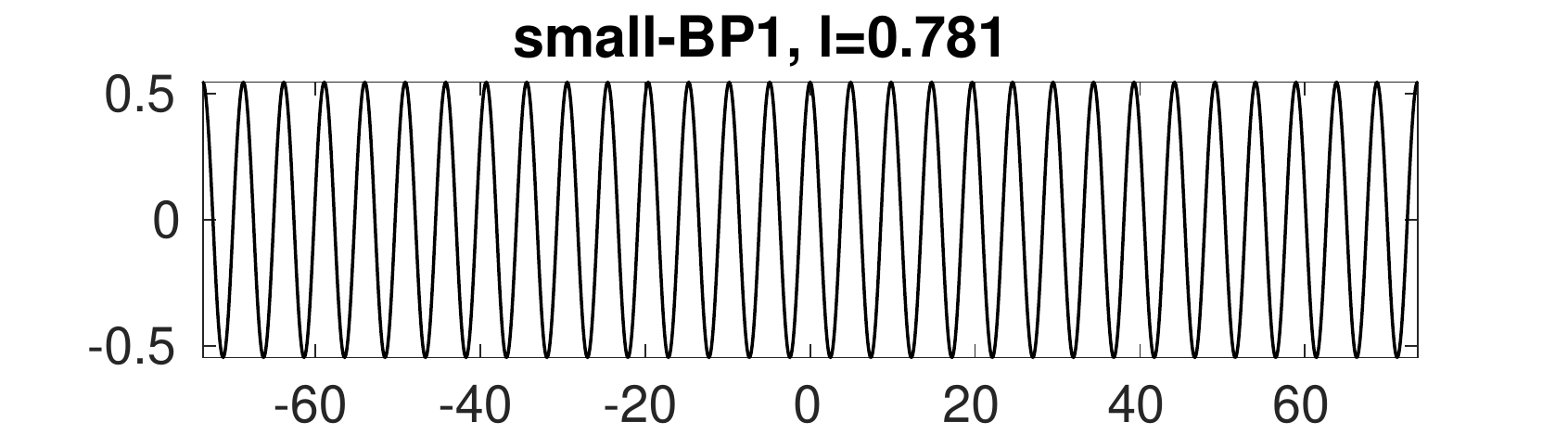}\\
\ig[width=0.4\twi,height=17mm]{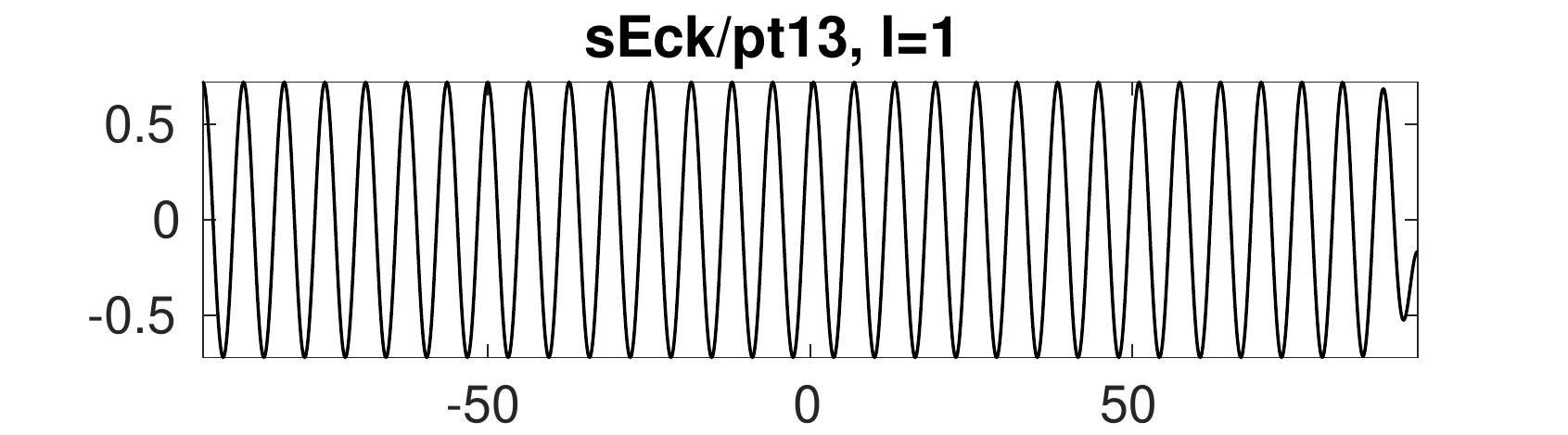}\\
\ig[width=0.4\twi,height=17mm]{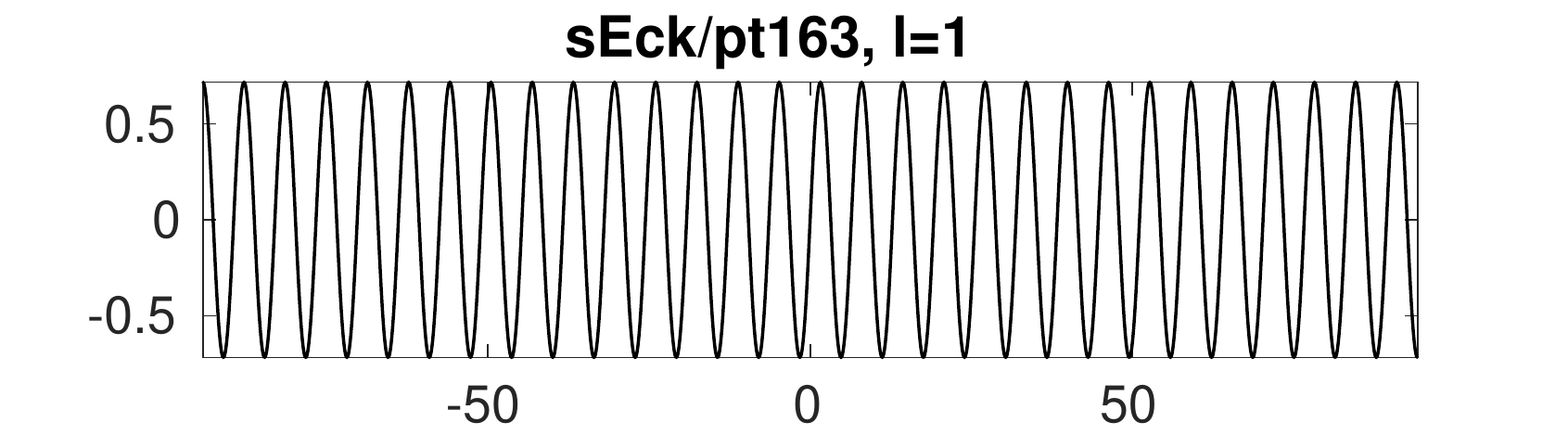}
\end{tabular}}
&\hs{-8mm}\raisebox{30mm}{
\begin{tabular}{l}
\ig[width=0.4\twi,height=17mm]{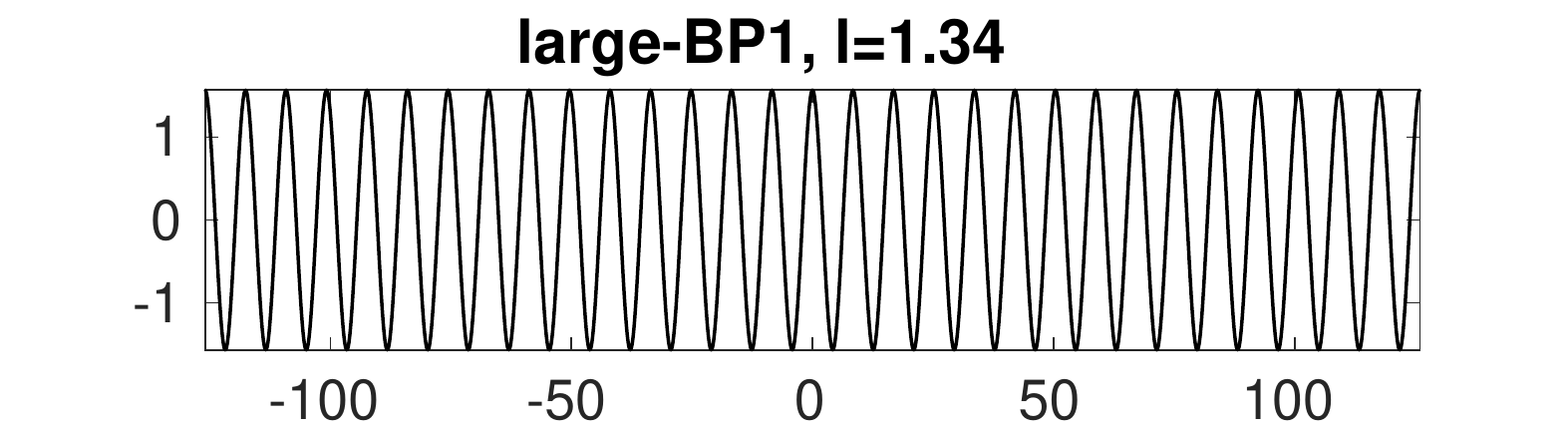}\\
\ig[width=0.4\twi,height=17mm]{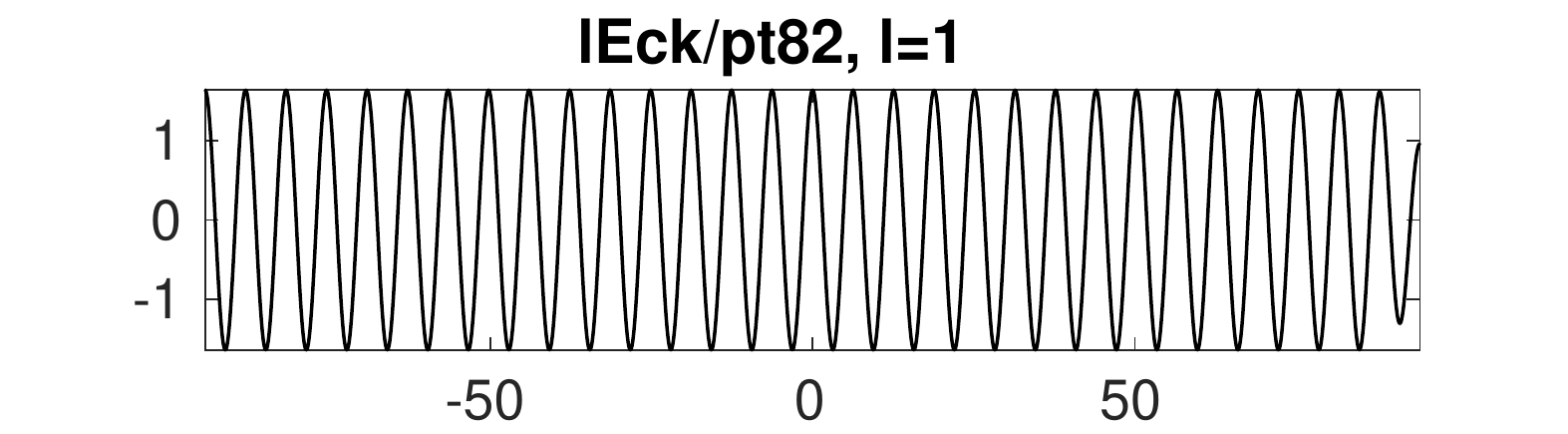}\\
\ig[width=0.4\twi,height=17mm]{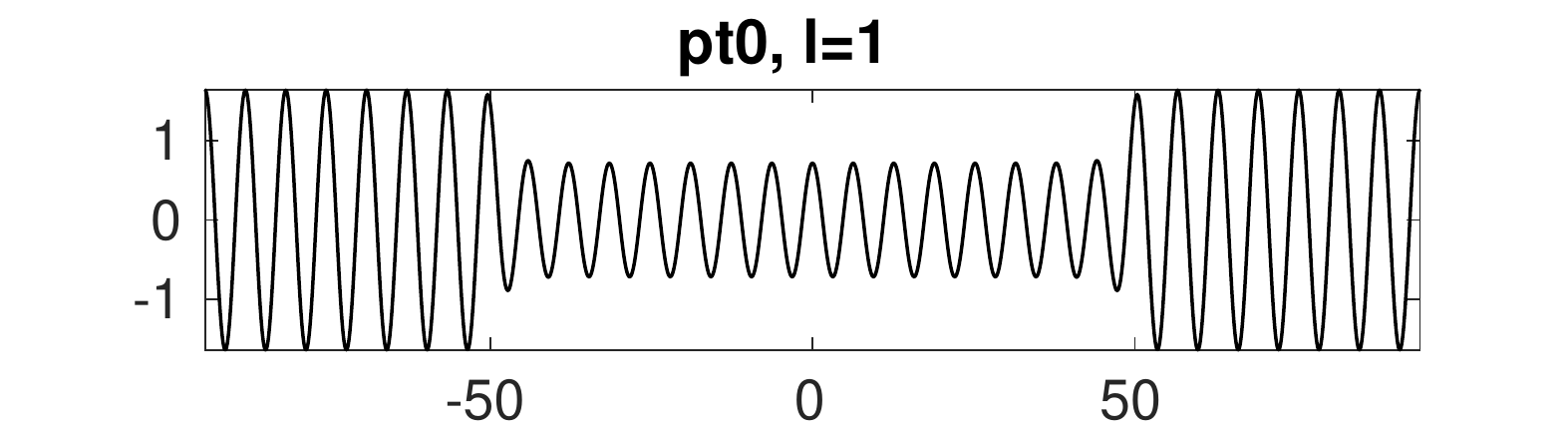}
\end{tabular}}
\end{tabular}\\[-1mm]
\ece
\vs{-2mm}
   \caption{{\small Continuation of solutions of \reff{sh1b} in $\ell$,  
with $a=5$, $\lam=1.3$ fixed on base domain $\Om=(-30\pi,30\pi)$, i.e.,
on the effective domain (also used for solution plots) $\Om_\ell=(-30\pi\ell,30\pi\ell)$.
Black branches correspond to $\upsg$ and $\uplg$, 
with Eckhaus BPs indicated and bifurcating branches shown in blue. 
See Fig.~\ref{fl2} for zoom and details of the red snake. \label{fl1}}}
\end{figure*}

\begin{figure*}[htbp]
\bce 
\begin{tabular}{lll}
(a)&(b)&(c)\\
\ig[width=0.25\twi,height=80mm]{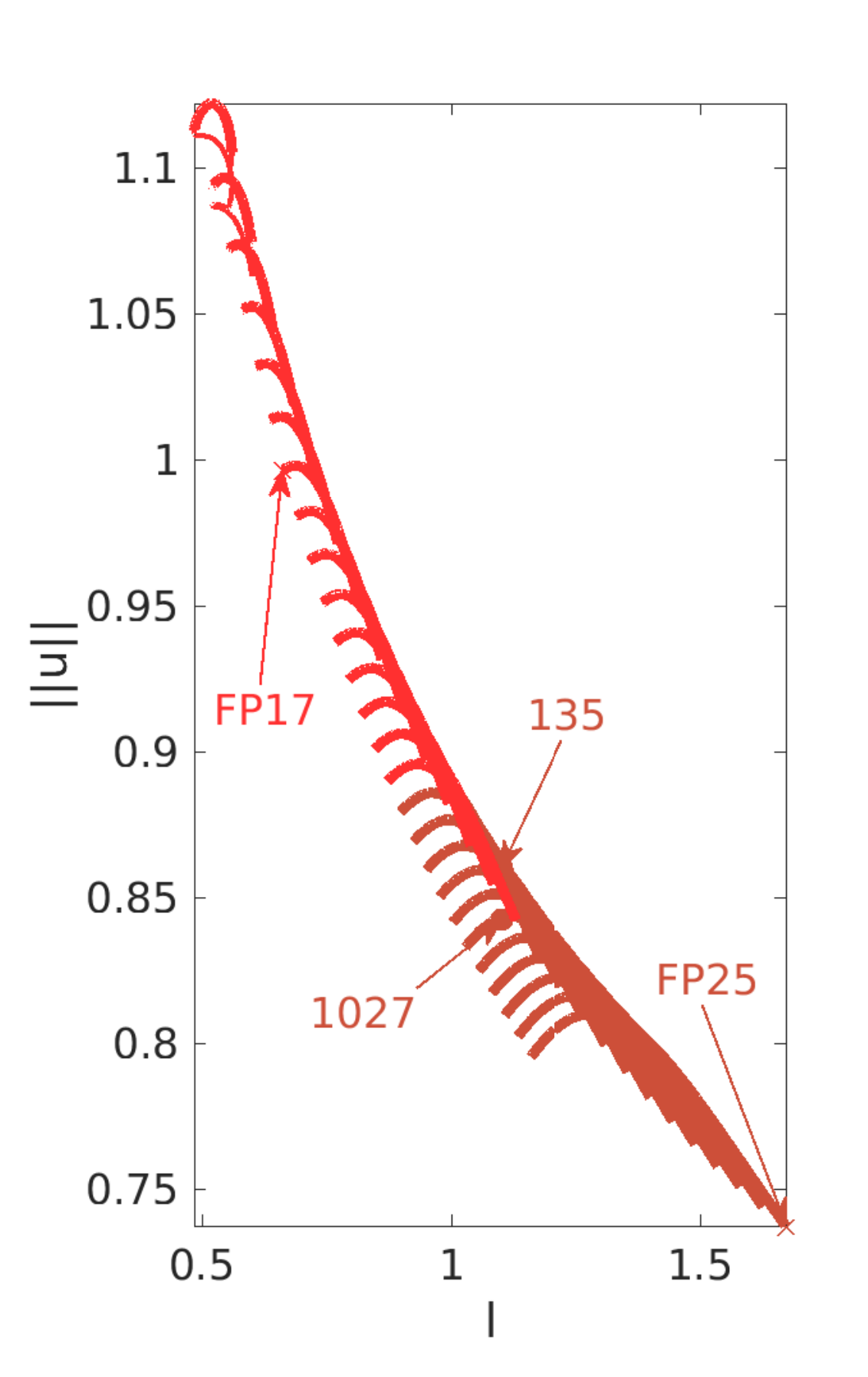}&
\hs{-3mm}\raisebox{40mm}{
\begin{tabular}{l}
\ig[width=0.4\twi,height=18mm]{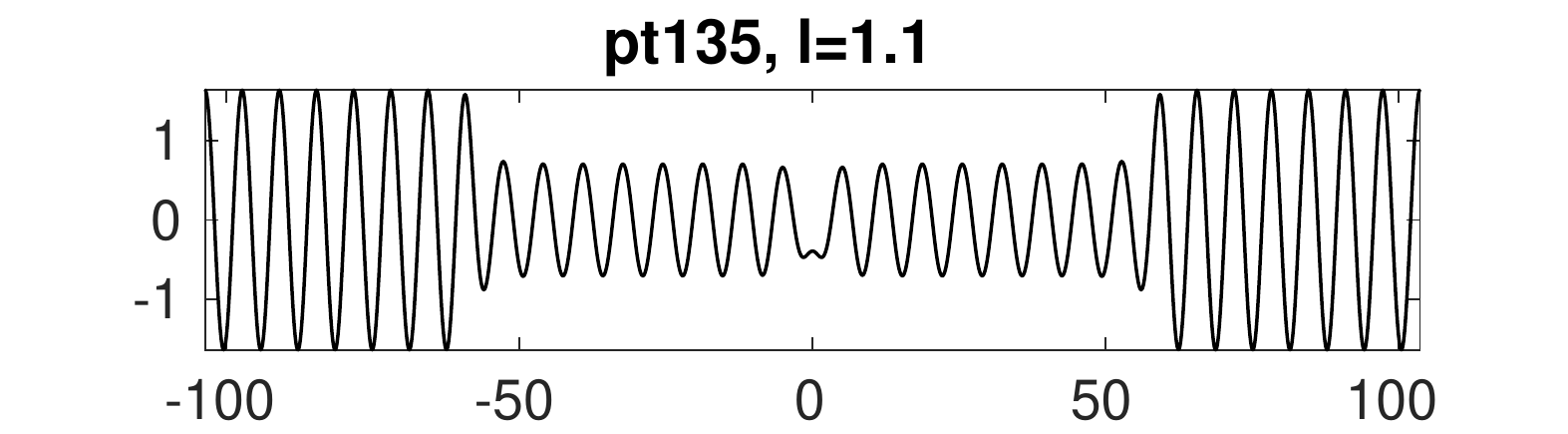}\\
\ig[width=0.4\twi,height=18mm]{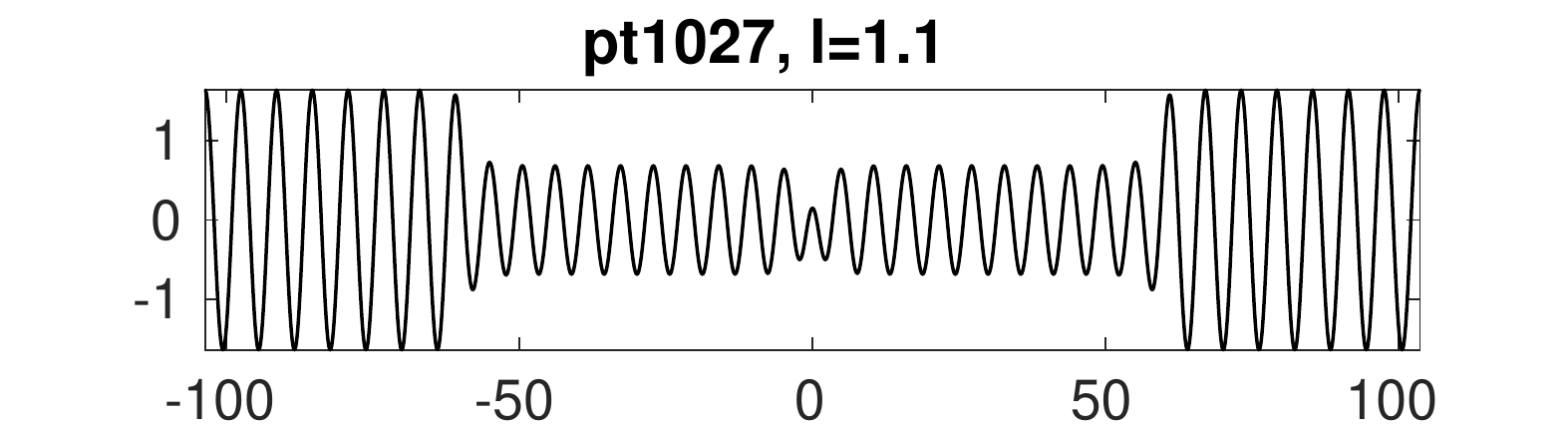}\\
\ig[width=0.4\twi,height=18mm]{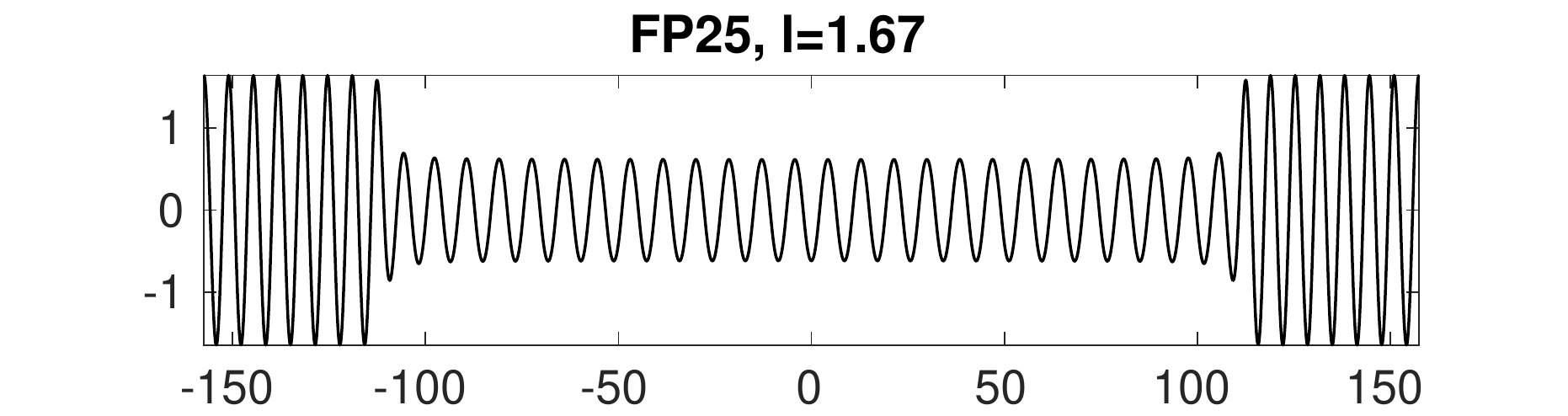}\\
\ig[width=0.4\twi,height=18mm]{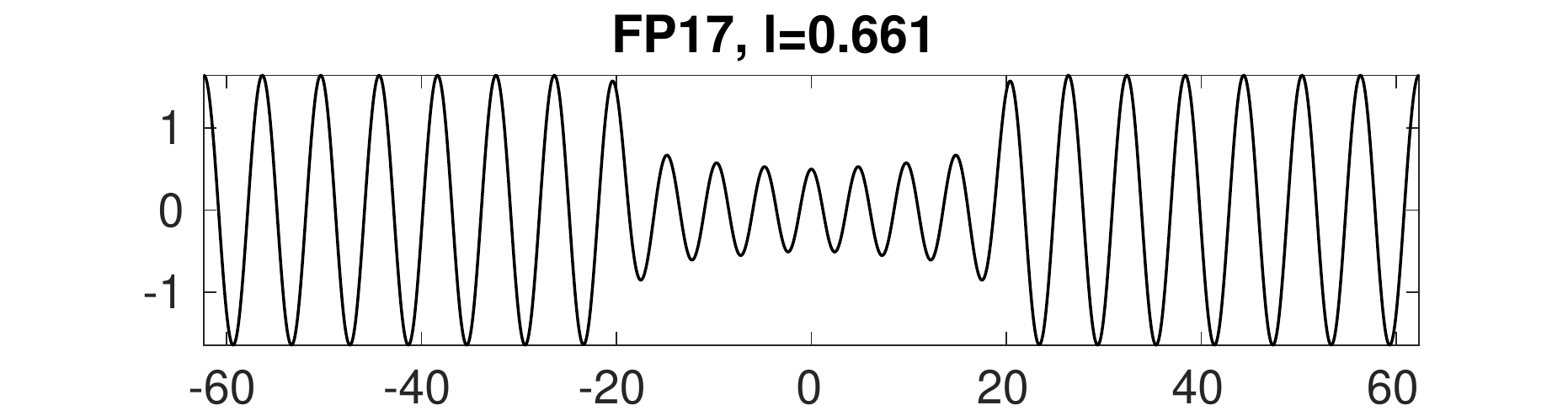}\\
\end{tabular}}&
\hs{-3mm}\raisebox{40mm}{
\begin{tabular}{l}
\ig[width=0.2\twi,height=19mm]{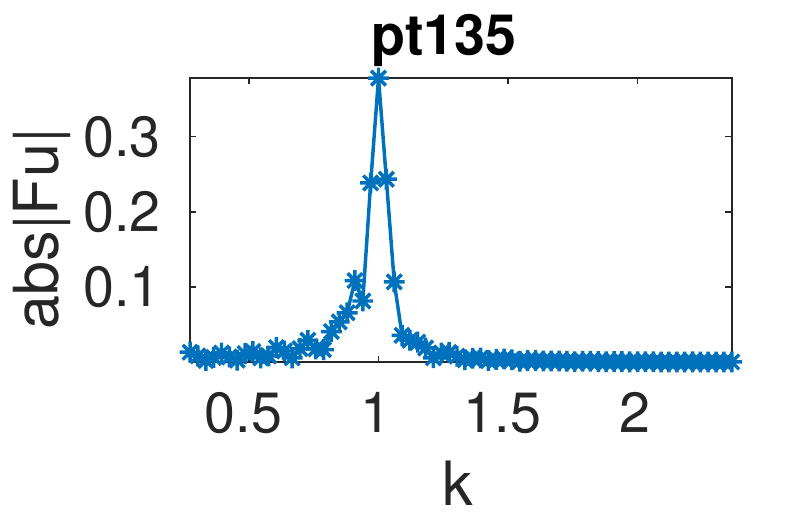}\\
\ig[width=0.2\twi,height=19mm]{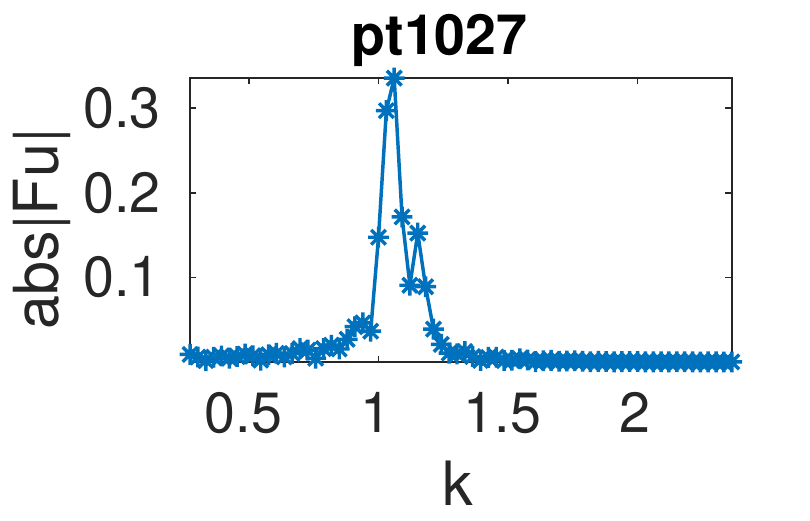}\\
\ig[width=0.2\twi,height=19mm]{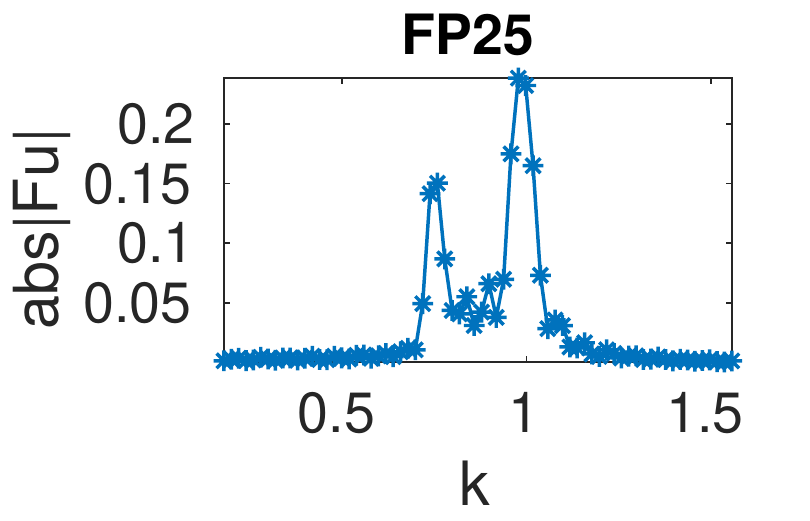}\\
\ig[width=0.2\twi,height=19mm]{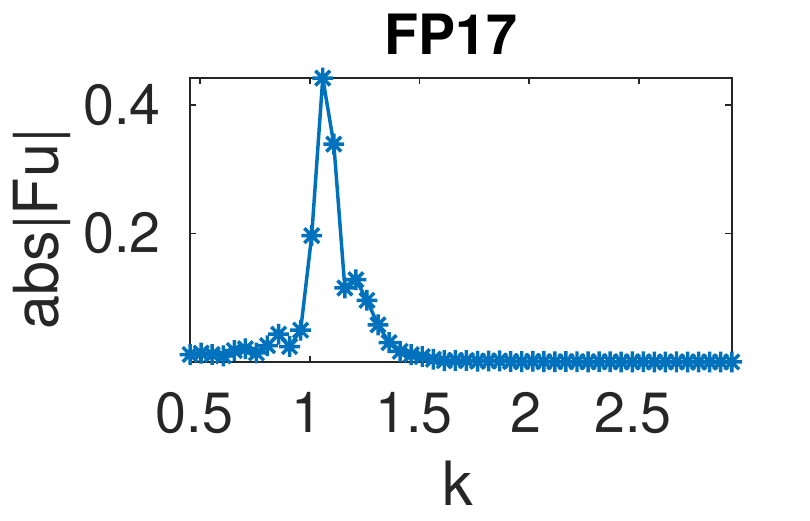}
\end{tabular}}
\end{tabular}\\
\begin{tabular}{lll}
(d)&(e)&(f)\\
\ig[width=0.25\twi,height=45mm]{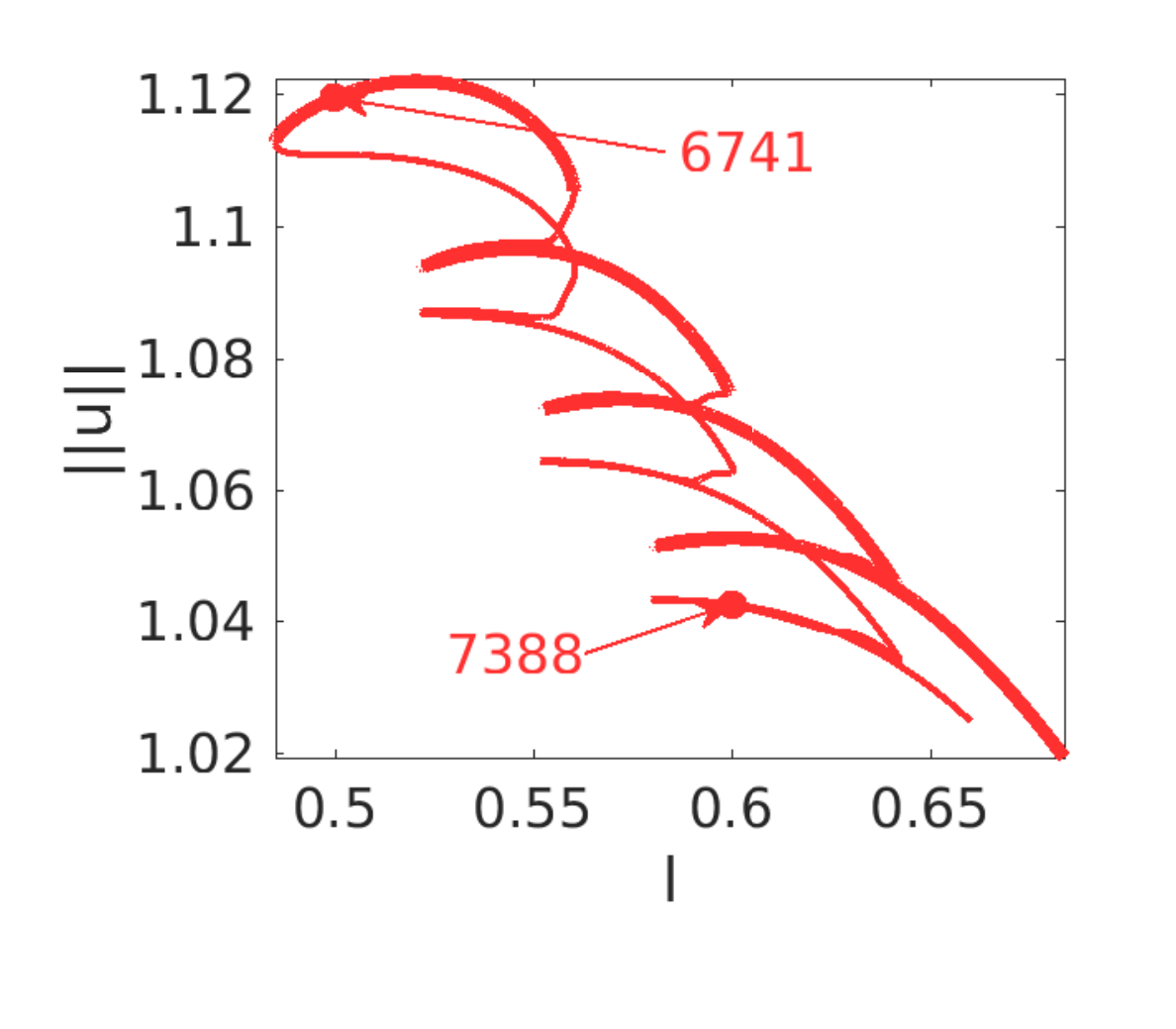}&
\hs{-3mm}\raisebox{25mm}{
\begin{tabular}{l}
\ig[width=0.40\twi,height=18mm]{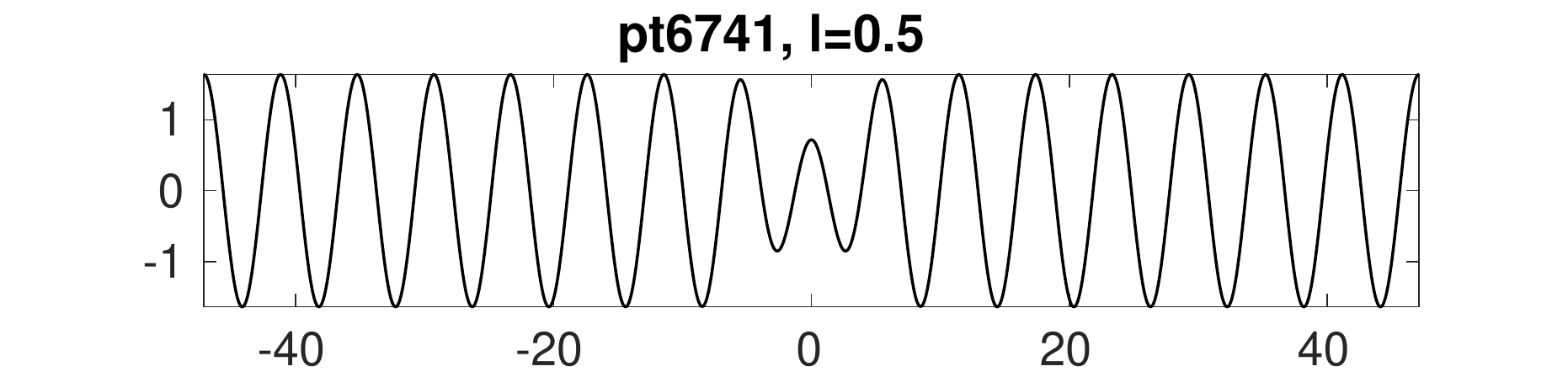}\\
\ig[width=0.40\twi,height=18mm]{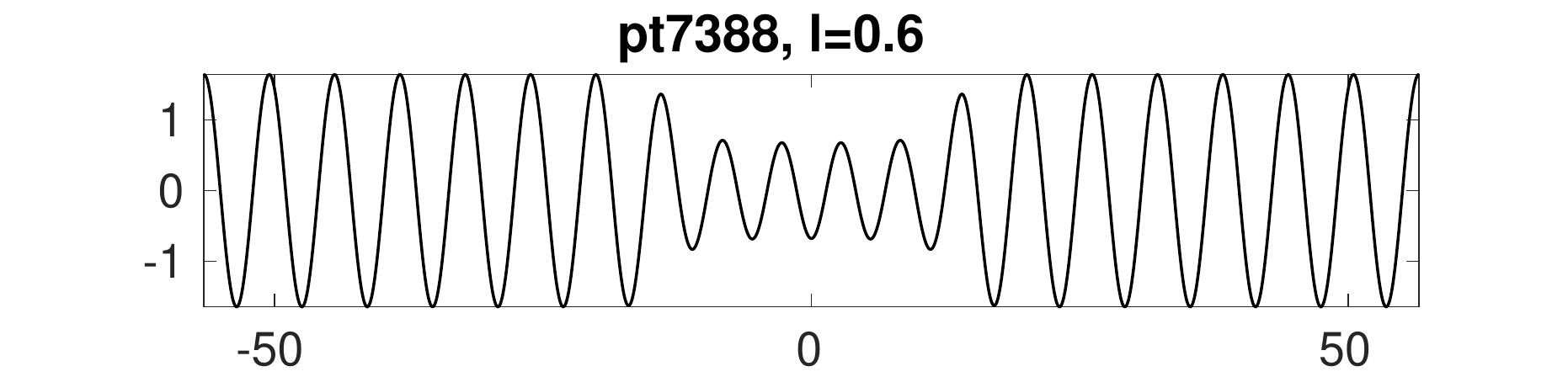}
\end{tabular}}&
\hs{-0mm}\raisebox{5mm}{\ig[width=0.23\twi]{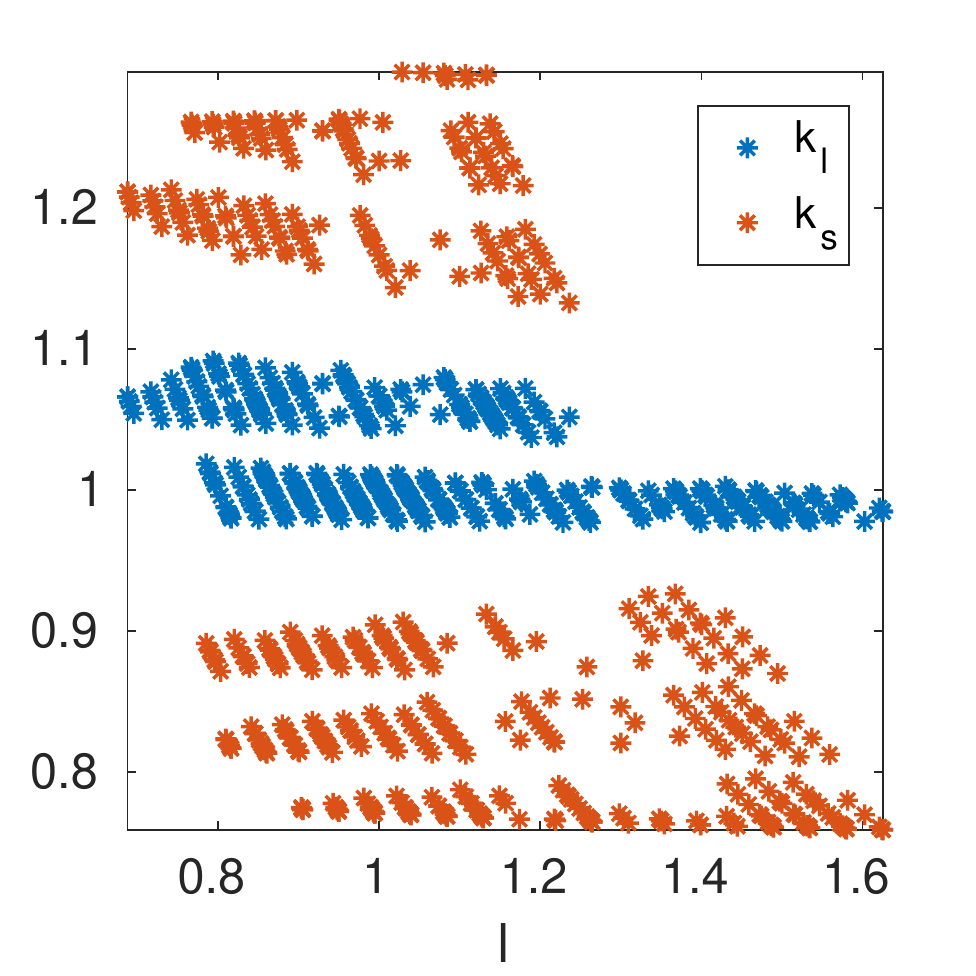}}
\end{tabular}
\ece
\vs{-7mm}
   \caption{{\small 
(a) Details of the $\uplg-\upsg-\uplg$ snake from Fig.~\ref{fl1}. (b,c) Sample solutions and Fourier transform.  (d,e) Left end of the 
snake and sample solutions. (f) Wave numbers of $\upsg$ (red) and $\uplg$ (blue) 
from the snake. \label{fl2}}}
\end{figure*}

In Fig.~\ref{fl1} we show the results of continuing the {\em periodic} 
branches $\uplg$ and $\upsg$ in $\ell$ for both increasing and decreasing $\ell$, starting
from results computed earlier for $\ell=1$ and $\lam=1.3$, and keeping $\lam$ fixed. For
decreasing $\ell$, $\upsg$ [lower black branch in Fig.~\ref{fl1}(a)] loses stability
at $\ell= \ell_0\approx 0.78$ corresponding to an Eckhaus boundary point for $\upsg$ with
wavelength $\ell_0$. The bifurcating branch {\tt sEck} initiates an amplitude 
modulation at the right domain boundary leading to an increase in the pattern
wavelength. With increasing $\ell$ this branch reconnects with the $\upsg$ branch
near $\ell=1.3$. In the displayed numerical continuation, this manifests itself
in 'branch jumping' to $\upsg$ with the corresponding $\ell$, which we deliberately do not
attempt to avoid here by refining the numerics. Instead, we simply continue the $\upsg$
branch so found back towards smaller $\ell$, and find that this branch, like the original
$\upsg$ branch, also loses stability near $\ell=0.8$. This process can then be repeated.
See (b) for sample solution profiles. We emphasize that the continuation of $\upsg$ in
$\ell$ maintains the number of wavelengths in the domain. Thus the number of wavelengths in
a solution can only change by encountering an Eckhaus point, i.e., by triggering a
(dynamic) phase slip that takes a stable periodic solution with a certain number of
wavelengths to a new stable periodic state with a different number of wavelengths \cite{kramerz85}.

Similar behavior takes place for $\uplg$ as well. With increasing $\ell$, $\uplg$ loses stability at an Eckhaus point at $\ell= \ell_0\approx 1.34$. Shortly thereafter the branch passes through a fold, and altogether we obtain a closed loop of $\uplg(\cdot;\ell)$ solutions [upper black branch in Fig.~\ref{fl1}(a)]. The bifurcation from $\uplg$ at $\ell_0$ again leads to amplitude modulation at the right domain boundary [Fig.~\ref{fl1}(c), middle], but in contrast to the $\upsg$ case, the bifurcating branch exhibits two folds near $\ell=1$ before reconnecting to $\uplg$ at the left Eckhaus boundary in $\ell$. See the first two plots in (c) for sample solutions. 

Additionally, Fig.~\ref{fl1}(a) shows a snaking branch (red) of 
$\uplg-\upsg-\uplg$ \hecy s, starting at $\ell=1$ ({\tt pt0}) with 
a sample profile shown on the bottom right of (c). This snaking in $\ell$ is 
shown in more detail in Fig.~\ref{fl2}. Starting at {\tt pt0} and 
increasing $\ell$ the solutions remain stable up until close to the 
first fold near $\ell=1.18$. Near this fold a phase slip is initiated.
This phase slip is once again associated with the onset of spatial modulation. This time
it is located in the middle of the domain and manifests itself in the splitting of the
central peak of the $\upsg$ portion of the profile [Fig.~\ref{fl2}(b)]. Beyond the fold
the new central peak regrows to the $\upsg$ amplitude but the solutions are now unstable,
and only recover stability at the next fold on the left. The net effect of this process
is to add half a wavelength of the $\upsg$ state to the solution profile. This process
then repeats adding a full wavelength of $\upsg$ in the middle of the 
domain after every two folds, and leading to slanted snaking with increasing $\ell$. It is clear that these repeated phase slips compress the $\upsg$ portion of the solution, although the wavelength of $\uplg$ also needs to adapt, albeit only slightly. See the first three plots in (b) for sample profiles. This process continues indefinitely.

If, on the other hand, we start at {\tt pt0} and decrease $\ell$, we can then extend the snake to the left, leading to a shrinkage of the middle section of $\upsg$ [see the last plot in (b) for a sample profile]. Panel (d) shows that this shrinkage is in all cases accompanied by significant hysteresis. Panels (d) and (e) show how the above process changes at the left end of the snake. At the leftmost fold [panel (d)], the amplitude of the last remaining $\upsg$ peak does {\em not} increase to the $\uplg$ amplitude. Instead the solution grows new $\upsg$ peaks on either side of the central peak thereby initiating a parallel snake similar to that just described, but with all solutions unstable.

To see the wave number adaption in a more quantitative way, in (c) we 
plot $|\hat u(k)|$ (the (discrete Fourier transform) of 
the sample solutions from (b). The larger peak, 
associated to $\uplg$, stays near $k=1$ throughout the snake, while the 
smaller peak, associated to $\upsg$ shifts. 
In (f) we summarize the wave numbers 
$k_s$ of the small peaks (red) and $k_l$ of the large peaks (blue) 
obtained from solutions in the snake. For this we only use solutions for which there is a clear peak separation in Fourier space, and thus solutions with $k_s\approx k_l$ are 
discarded. The plot shows that $k_l$ adapts much less than $k_s$. While some scatter due to the limited resolution in Fourier space is present, bands in the wave numbers corresponding to different forward and backward transitions in the snake are clearly visible. Because of the ability of both $\upsg$ and $\uplg$ states to absorb the wavelength change generated by the phase slips in the center of the domain we conjecture that this type of double structure is more robust with respect to wavelength change than structures that are more rigid. 

The above scenario is reminiscent of defect-mediated snaking \cite{Ma}, with two important differences. In defect-mediated snaking the solution branch appears from a fold on the branch of homogeneous states and so consists of a single branch. Moreover, the snaking is not slanted, because the phase slips do not result in the compression of a second portion of the solution. It is this compression that is in our case responsible for the slanted nature of Fig.~\ref{fl2}(a) since it turns the problem into an effectively nonlocal one.

\subsection{The case $a=9.5$: breakup of snakes into stacks of isolas, and a multitude of cycles} \label{ss95}

\begin{figure}[htpb]
\bce 
\begin{tabular}{lll}
(a)&(b)&(c)\\
\hs{-2mm}\ig[width=0.16\twi,height=58mm]{h9-5}&
\hs{-2mm}\ig[width=0.12\twi,height=58mm]{l2-9-5}&
\hs{-1mm}\ig[width=0.19\twi,height=60mm]{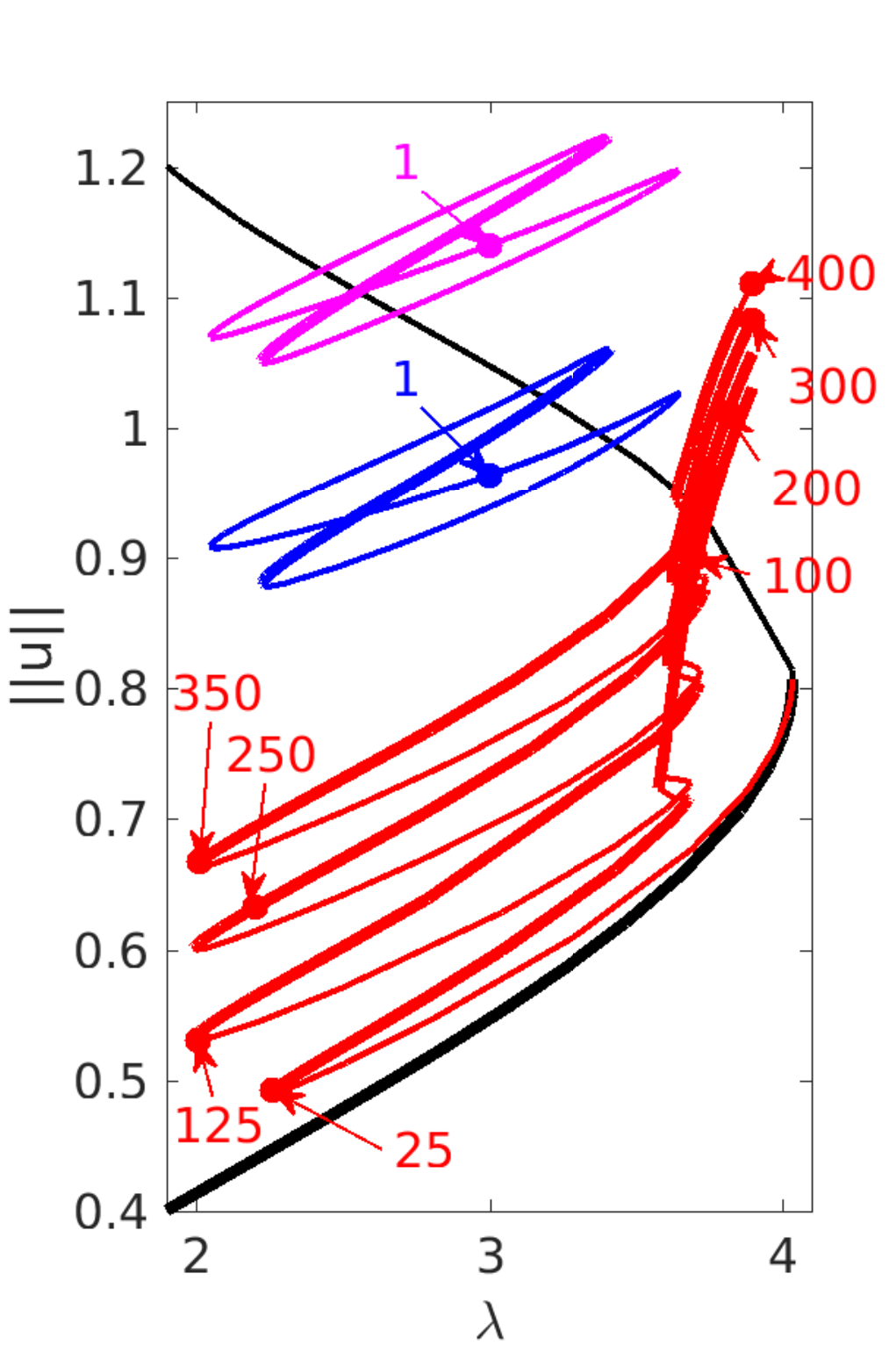}
\end{tabular}\\
\begin{tabular}{l}
(d)\\
\ig[width=0.22\twi,height=14mm]{n25}\ig[width=0.22\twi,height=14mm]{n100}\\
\ig[width=0.22\twi,height=14mm]{n125}\ig[width=0.22\twi,height=14mm]{n200}\\
\ig[width=0.22\twi,height=14mm]{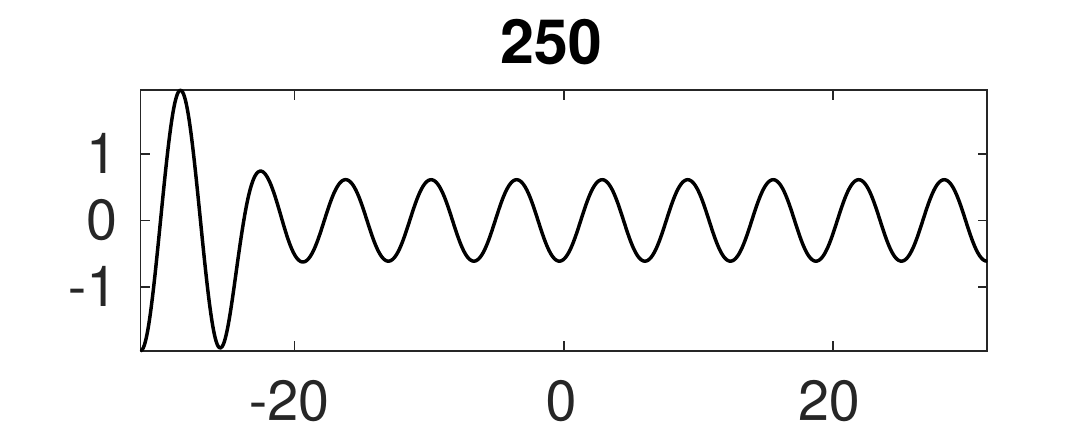}\ig[width=0.22\twi,height=14mm]{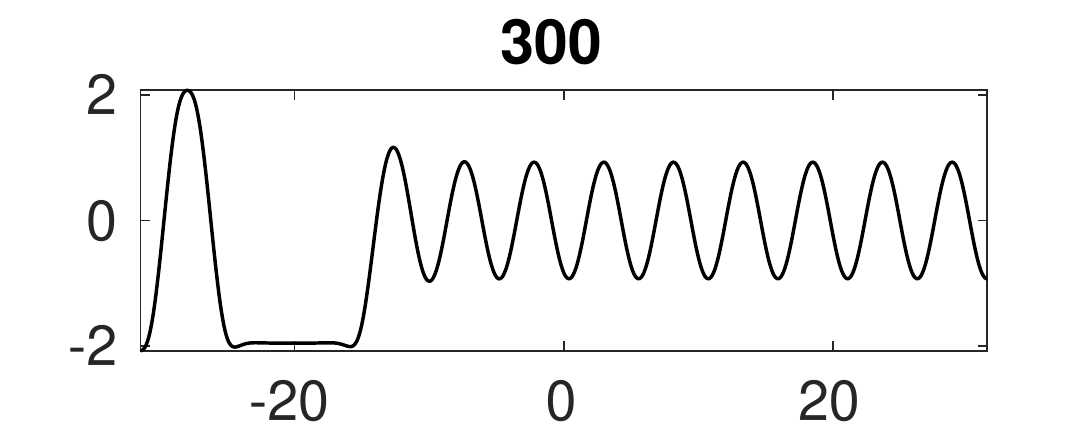}\\
\ig[width=0.22\twi,height=14mm]{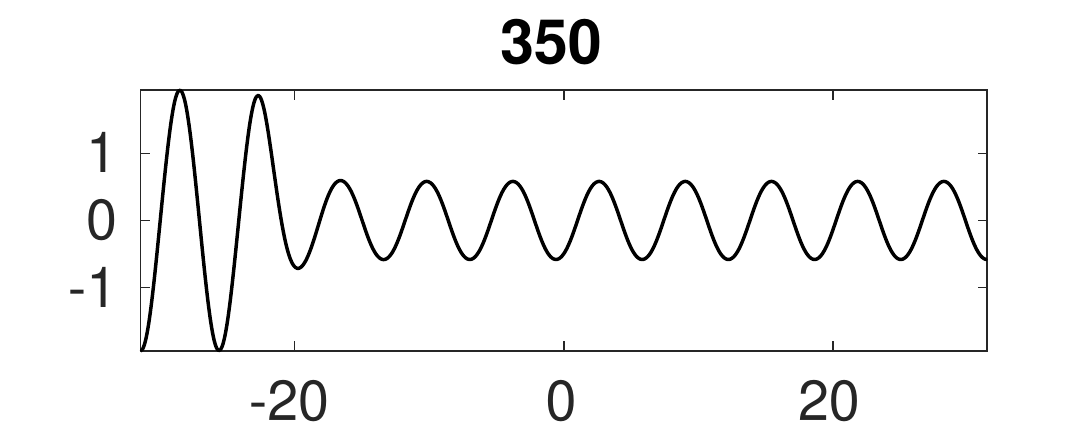}\ig[width=0.22\twi,height=14mm]{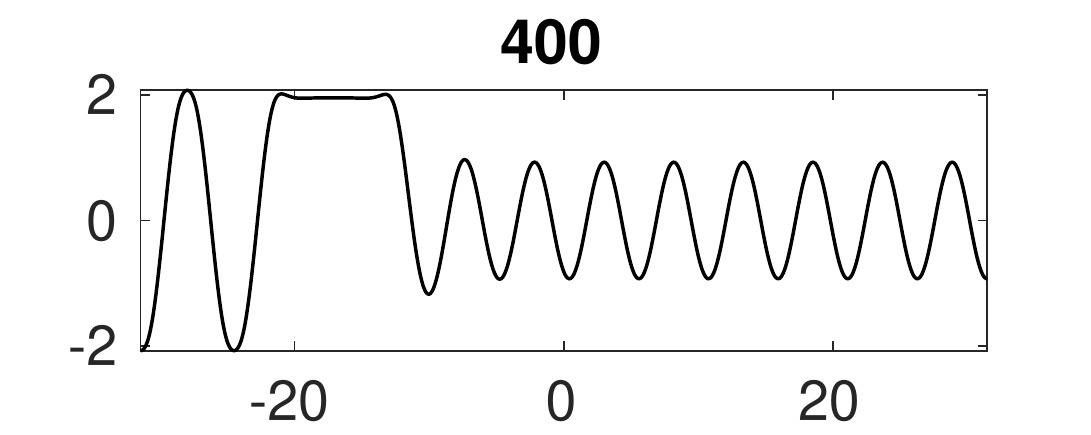}
\end{tabular}
\ece
\vs{-2mm}
   \caption{{\small 
$a{=}9.5$, $\Om{=}(-10\pi,10\pi)$. (a) $H$ for solution branches plotted as in Fig.~\ref{f1}(a); there are now multiple intersections near $\lam=3.5$. (b) 
$\|u\|$ for the $k{=}1$ and $k{=}0$ branches, illustrating the multistability near $\lam=3.5$. (c) Two isolas of cycles between $\ups$ and 
$\upl$ obtained from suitable initial guesses, further discussed in Fig.~\ref{f3b}, and a branch (red) of 
$\upl$--$\ups$ cycles which passes near $\lam{=}3.5$ close to $\pm\uhl$. 
(d) Sample solutions from the red branch in (c).  \label{f3}}}
\end{figure}

Larger values of $a$ result in more and more intersections among $H$ 
for periodic branches with different $k$, and the homogeneous branch $k=0$
[see Fig.~\ref{f3}(a) for $a=9.5$], and thus many more \hecy s become possible. 
Up to $a=9.3$ (say, and depending on the domain size), the basic 
snake--and--ladders structure from Fig.~\ref{f1} stays intact, but for larger 
$a$ the snakes and rungs reconnect into a stack 
of isolas. Moreover, $\uhl$ plays an increasing role in the continuation 
of the solutions. Figures~\ref{f3}(c)-(d) illustrate these effects. The  
red branch bifurcates from BP1 on $\up$, but fails to 
grow a front between $\upl$ and $\ups$ as it would in 'classical' snaking 
at lower $a$. Instead, it 
now exhibits long, nearly vertical intervals located near $\lam=3.5$, associated
first with the growth of a segment of $-\uhl$ in the solution profile and then its
shrinkage before another half period of $\uplg$ can be inserted. The branch bifurcating
at BP2 on $\upsg$, which at lower $a$ generates a snaking branch of
$\upsg-\uplg-\upsg$ \hecy s, cf.~Fig.~\ref{f1}, behaves similarly and also always involves plateaus of $\pm\uhl$.

This type of behavior is similar to that recently found in a Gray-Scott model studied
in connection with dryland vegetation patterns where tristability between a pair of
different spatially periodic states and a homogeneous state is also present
\cite{Gandhi2018}, suggesting that the behavior shown in Fig.~\ref{f3}(c) is in fact
generic. 

\begin{figure}[htpb]
\bce 
\begin{tabular}{l}
(a)\\
\ig[width=0.25\twi,height=50mm]{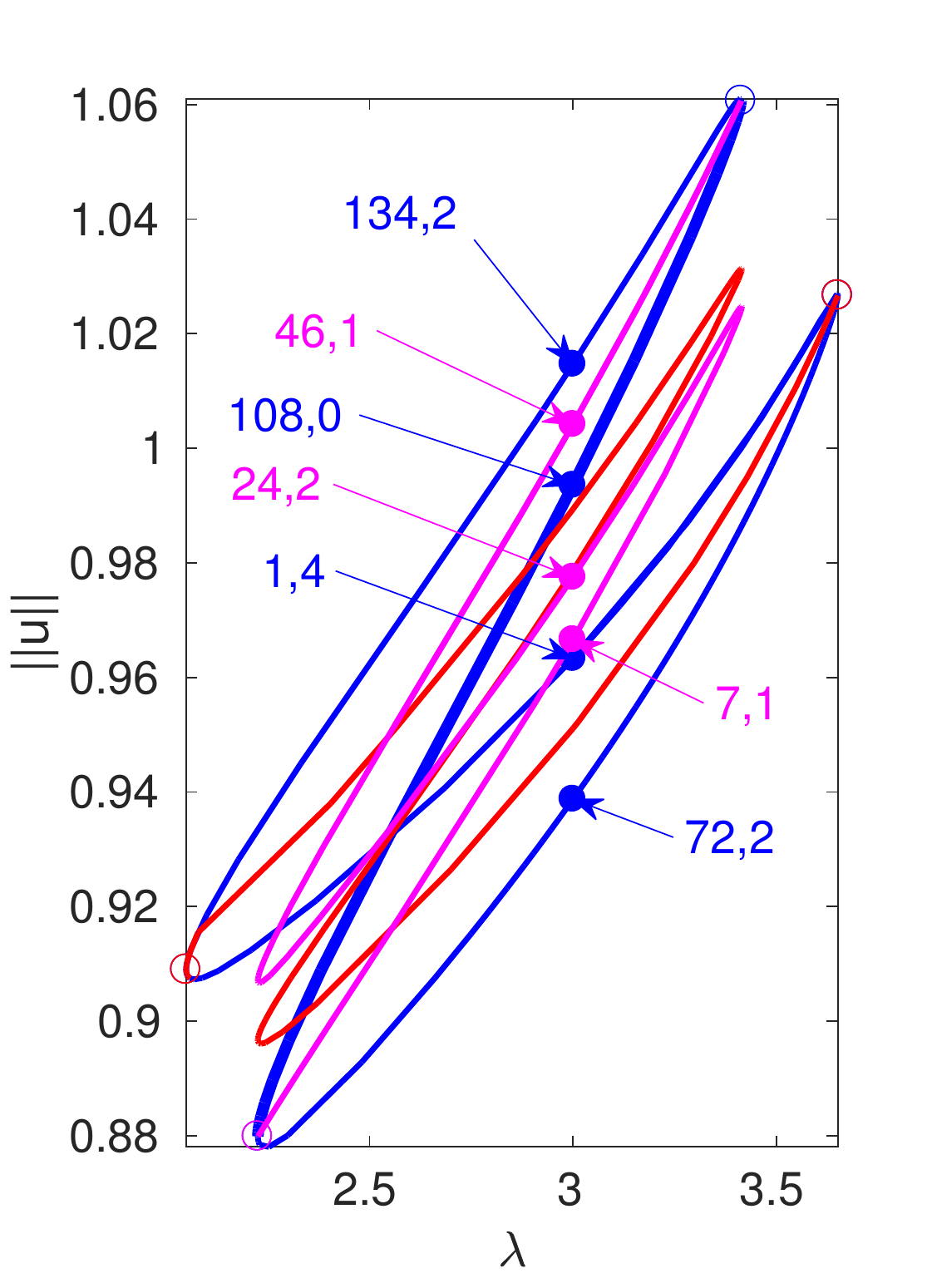}\\
(b)\\
\hs{-0mm}\raisebox{0mm}{\begin{tabular}{l}
\ig[width=0.21\twi,height=14mm]{9-li1-1}\ig[width=0.21\twi,height=14mm]{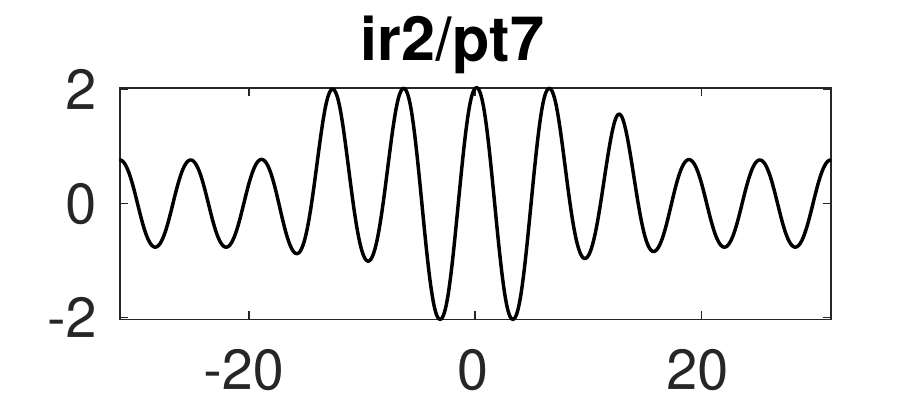}\\
\ig[width=0.21\twi,height=14mm]{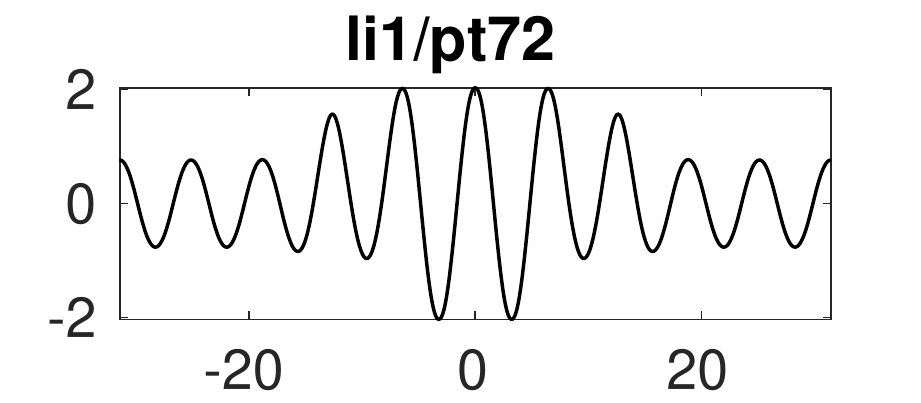}\ig[width=0.21\twi,height=14mm]{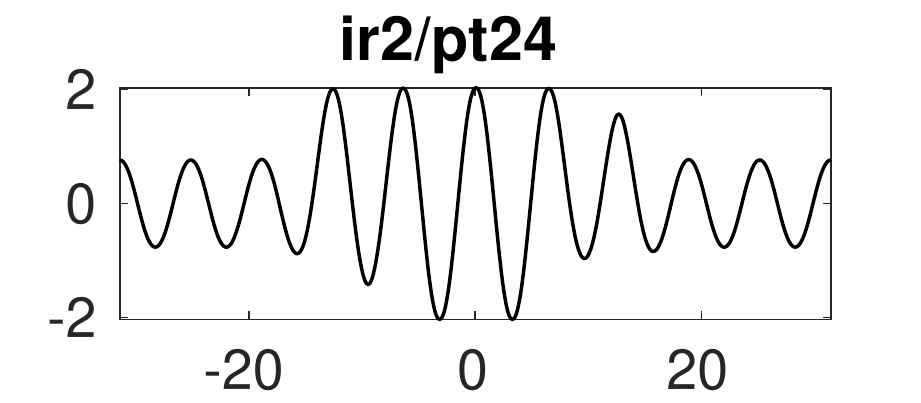}\\
\ig[width=0.21\twi,height=14mm]{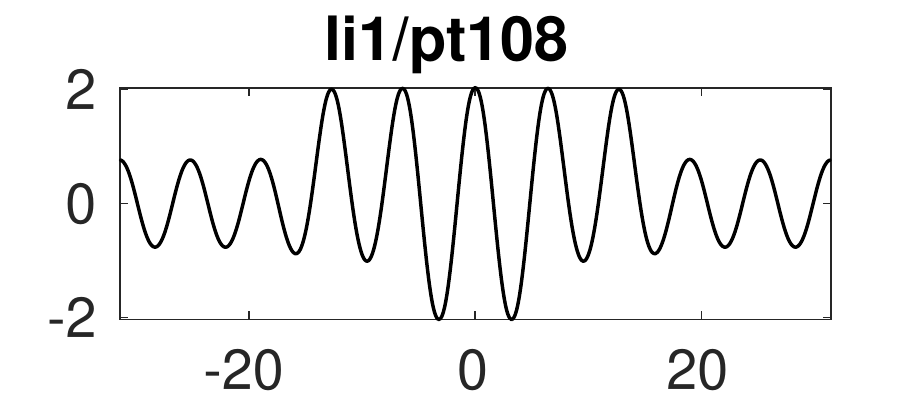}\ig[width=0.21\twi,height=14mm]{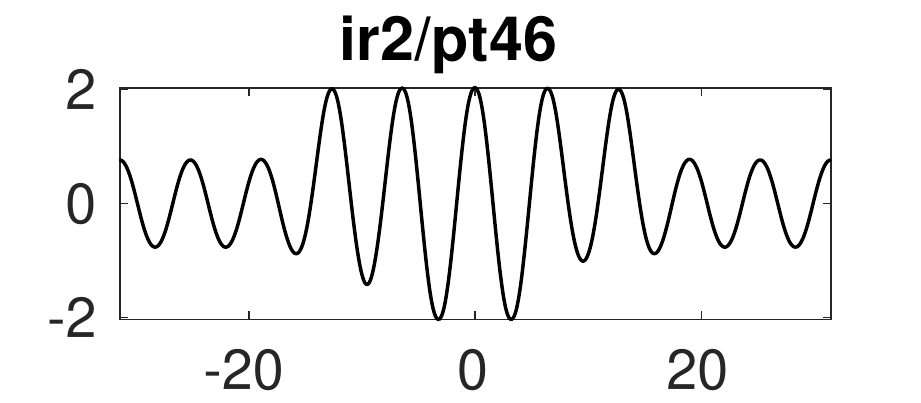}\\
\ig[width=0.21\twi,height=14mm]{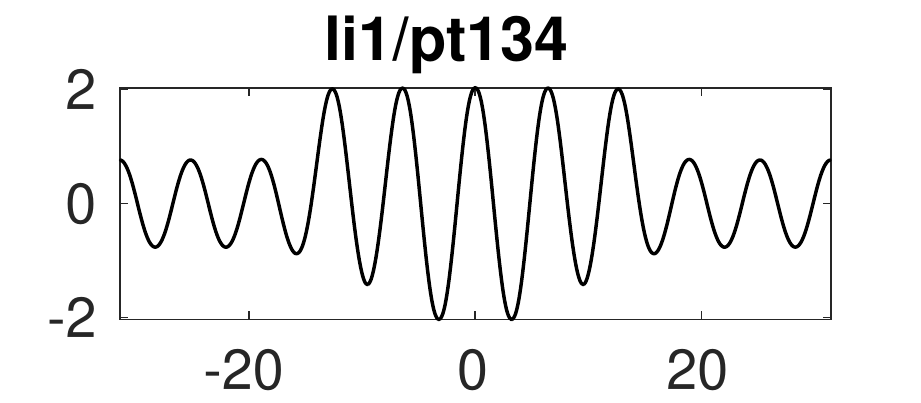}\ig[width=0.21\twi,height=14mm]{9-li2-1}
\end{tabular}}
\end{tabular}
\ece
\vs{-2mm}
   \caption{{\small  (a) Zoom of the 
blue isola in Fig.~\ref{f3}(c) and associated rung structure. The labels ptnr,$i$ 
indicate stability properties with $i$ specifying the number of unstable
eigenvalues. (b) Sample solutions from (a).  \label{f3b}}}
\end{figure}

On the other hand, we can still generate 'classical' \hecy s between $\upsg$ and $\uplg$ 
leading to, e.g., the blue and magenta isolas in Fig.~\ref{f3}(c). To generate starting
points for these we glue together $\ups$, $\upl$ segments (using a longer middle segment 
for the magenta isola) and running a Newton loop to converge to a solution.
Figure \ref{f3b} shows details of the reorganization of the segments 
that make up the snakes and rungs at smaller $a$ into the isola structure.

\begin{figure*}[htpb]
\bce 
\begin{tabular}{ll}
{\small (a)}&{\small (b)}\\
\ig[width=0.18\twi,height=28mm]{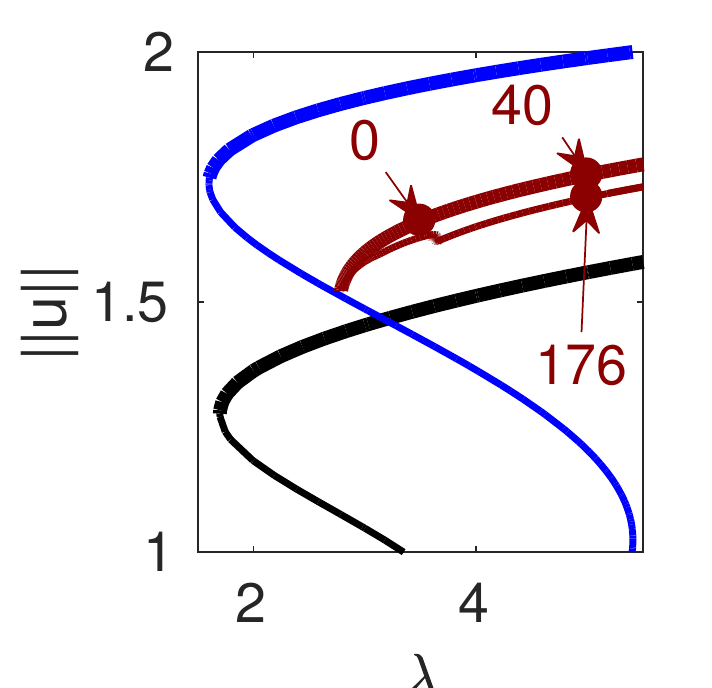}&
\raisebox{4mm}{
\ig[width=0.18\twi,height=24mm]{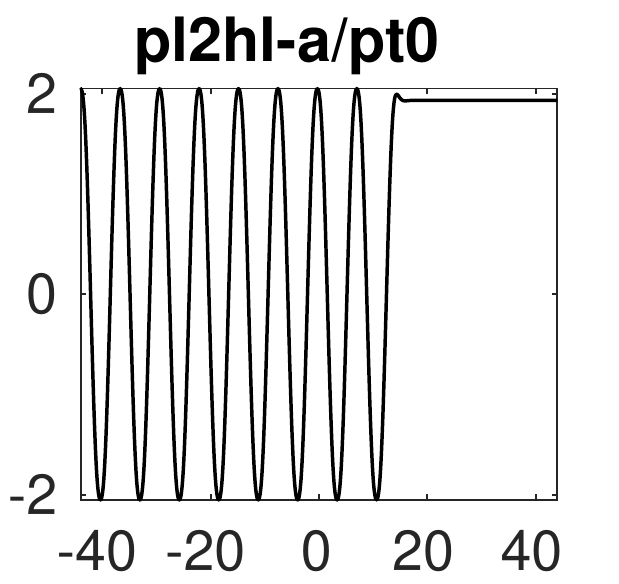}
\ig[width=0.18\twi,height=24mm]{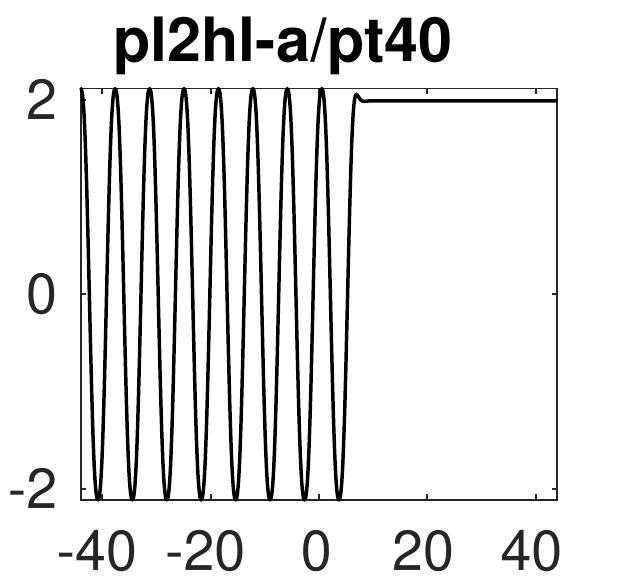}\ig[width=0.18\twi,height=24mm]{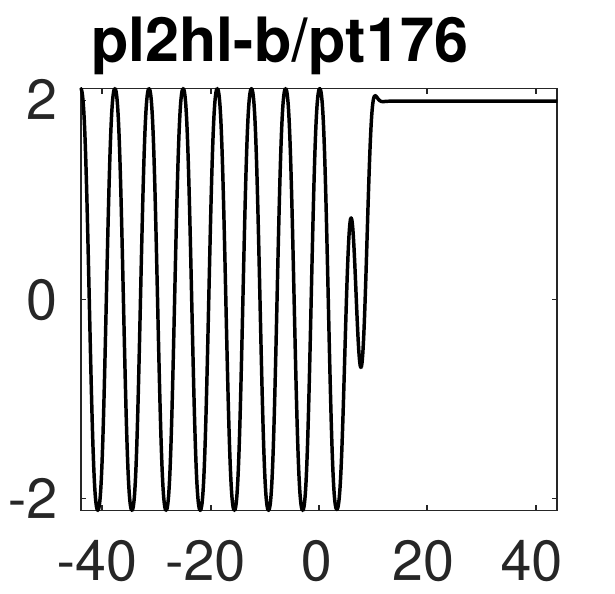}}\\
{\small (c)}&{\small (d)}\\
\ig[width=0.23\twi,height=40mm]{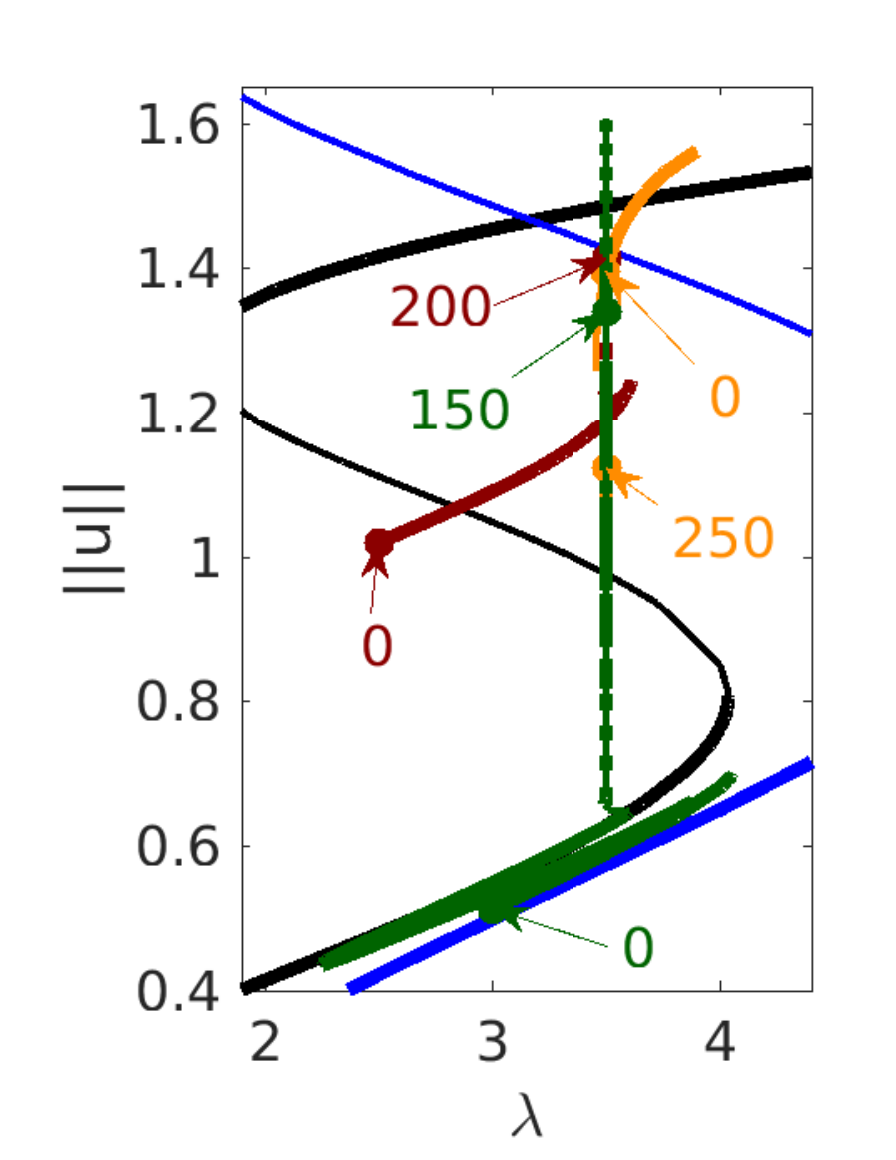}&
\raisebox{20mm}{\begin{tabular}{l}
\ig[width=0.18\twi,height=20mm]{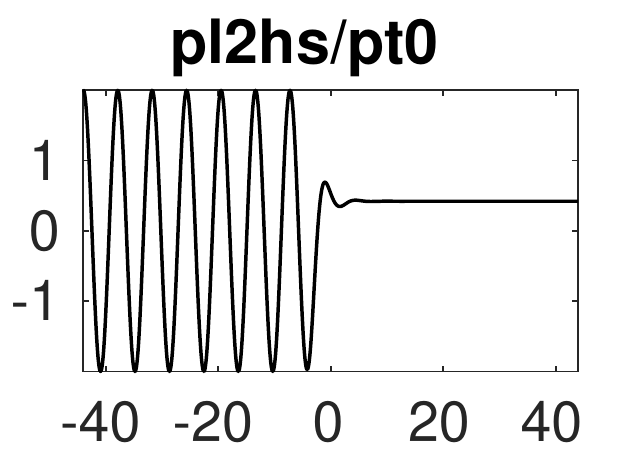}
\ig[width=0.18\twi,height=20mm]{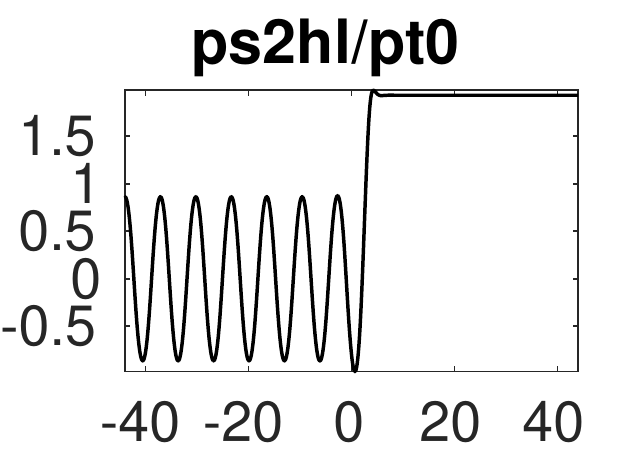}\ig[width=0.18\twi,height=20mm]{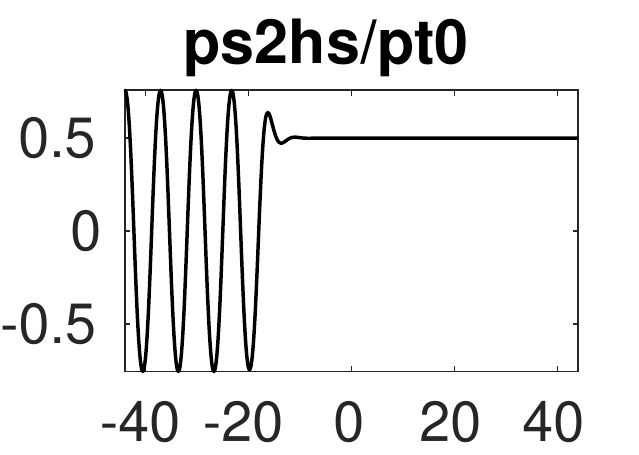}\\
\ig[width=0.18\twi,height=20mm]{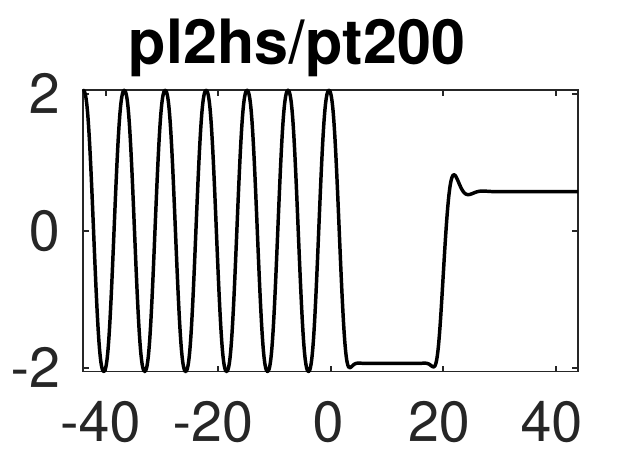}
\ig[width=0.18\twi,height=20mm]{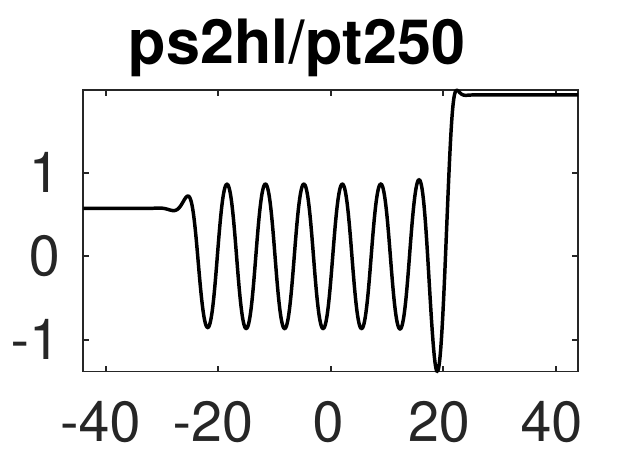}\ig[width=0.18\twi,height=20mm]{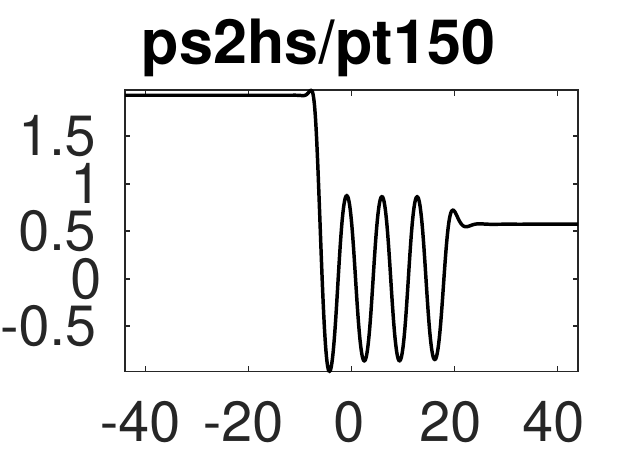}
\end{tabular}}
\end{tabular}
\ece
\vs{-6mm}
   \caption{{\small Various localized patterns at $a=9.5$, $\Om=(-14\pi,14\pi)$, 
with starting poinnts for the continuation obtained from running a Newton loop on 
rough initial guesses, for instance 
gluing together a large pattern segment and the large homogeneous solution. (a,b) Bifurcation diagram and sample plots 
from  the red branch {\tt pl2hl} 
('pattern - large - to - homogeneous - large') in (a). 
(c) Bifurcation diagram of branches  {\tt pl2hs} (red), {\tt ps2hl} (green) and 
 {\tt ps2hl} (orange). (d) Sample solutions from (c). 
 \label{f4}}}
\end{figure*}

Figure~\ref{f3}(d) suggests that for $a=9.5$ we can expect further cycles, 
not present at small $a$, where the $\uhl$ state plays a prominent role. 
Figure~\ref{f4} provides some examples. The red branch in (a), with sample solutions
in (b), corresponds to $\uplg-\uhl$ heteroclinics. In states of this type it is
difficult to eliminate a wavelength of the periodic state in response to
parameter changes -- it costs less energy to compress the state and expand
the homogeneous state or vice versa. Thus phase slips will not be triggered
unless the wavelength of the periodic state changes by a substantial amount. 
Solutions on the lower part of this branch (with {\tt pt176}) include short loops near $\upsg$ and are unstable. 
Panels (c) and (d) show more \hecy s, indicating that at large $a$ (and on sufficiently
large domains) all sorts of connections are possible provided one selects $\lam$ values
corresponding to intersections of $H$ for the pertinent patterns. We used initial guesses of the form 'pattern-large-to-homogeneous-small' (pl2hs, red), 'pattern-small-to-homogeneous-small' (ps2hs, green), 'pattern-small-to-homogeneous-large' (ps2hl, orange) to converge to the corresponding solutions and continued the resulting solutions to $\lam\approx3.5$, where they all form localized patterns consisting of (on this domain) up to 4 patches of different solutions ($\uhs, \uhl, \upsg$, and $\uplg$), many of which are stable, albeit in rather narrow $\lam$ intervals. These stability intervals are a consequence of what appears to be collapsed snaking of the corresponding branches in the vicinity of the Maxwell point at $\lam\approx3.5$. 
In states of this type changes in the wavelength
of the stripe portion of the solution are readily accommodated by changes in the
homogeneous portion. 

The reconnection of the snaking diagram into a stack of isolas as parameters are varied
has been seen in SH23 \cite{burke,BKLS09} and in two-dimensional patterns may even occur
as one proceeds up a snaking diagram, all other parameters remaining fixed
\cite{msand10,bramburger}. 

\section{Discussion}\label{dsec}

In suitable parameter regimes, the SH357 equation \reff{sh2} allows many different 
\hecy s between four (recall that we identify $\pm \upsg, \pm\uplg, 
\pm \uhs$ and $\pm\uhl$, respectively) main building blocks, namely  
\bcen 
\item $\upsg$ and $\uplg$: periodic patterns of small and large amplitude, 
respectively, with wave numbers $k$ near $1$. 
\item $\uhs$ and $\uhl$: spatially homogeneous states of small and large amplitude; these can be seen as special cases of $\upsg$ and $\uplg$ with 
wave number $k=0$. However, from the point of view of \hecy s, 
the homogeneous states are special in the sense that segments 
of $\uhs$ and $\uhl$ can have arbitrary length. 
\ecen 
To give some structure to our results, we made the special choice 
\reff{bsel}, i.e., $b=3.5+0.4(a-3)$. For relatively small $a$ ($a=2$ in \S\ref{ss2}),
we have a more or less simple situation in the sense that the necessary condition 
\huga{\label{Hd}
H(u_1(\cdot;\lam))=H(u_2(\cdot;\lam))
}
for the patterns involved in a \hecy\ only holds for a relatively 
small number of patterns. Moreover, these values of $a$ allow for an interesting 
continuous transition between \hecy s between $\uplg$ and 
$\upsg$, and \hecy s between $\uplg$ and $u\equiv 0$. For intermediate 
$a$ ($a=5$ in \S\ref{ss5}), we obtain 'classical' snaking of \hecy s 
between $\uplg$ and $\upsg$. Via continuation in the domain size we also 
found a new kind of slanted snaking (reminiscent of 
defect-mediated snaking), where, e.g., the $\upsg$ portion of the solution
grows or shrinks via phase slips. Finally, for 'large' $a$, 
the $\upsg$--$\uplg$--$\upsg$ snaking breaks up into stacks 
of isolas. Moreover, \reff{Hd} is fulfilled for a large number of 
different patterns, and consequently many different \hecy s become possible. 
In \S\ref{ss95}, we provide several examples of such states for $a=9.5$. 

In our bifurcation diagrams we indicated linearly stable (unstable) solutions 
by thick (thin) lines. Owing to the generally large number of different simultaneously
stable solutions, periodic and localized, it would be desirable to characterize in
addition the different basins of attraction. However, such basins  
are not yet well understood even for standard homoclinic snaking as described 
by SH23 or SH35. Therefore, here we confine ourselves to a few remarks: 
The snakes and related structures such as stacks of isolas 
typically have rather large basins. For instance 
the starting points for the computation of the isolas in Fig.~\ref{f3} 
can also be obtained from quite rough initial guesses
followed by time integration towards a stable solution. 
Of course, different
initial guesses may lead to different but 'nearby' steady states. Put differently, 
a localized $O(1)$ perturbation of a solution $u^*$ in a snake typically leads to 
rather fast relaxation either back to $u^*$, or to a nearby solution in the snake, 
with, say, one more 'large' roll and one fewer 'small' roll. This behavior naturally
changes if we go outside the snaking region, or consider delocalized perturbations, 
which in particular may lead to depinning, as illustrated next. 

\begin{figure*}[htpb]
\bce 
\begin{tabular}{lll}
(a1) &(b) &(d) \\ 
\ig[width=0.32\twi,height=20mm]{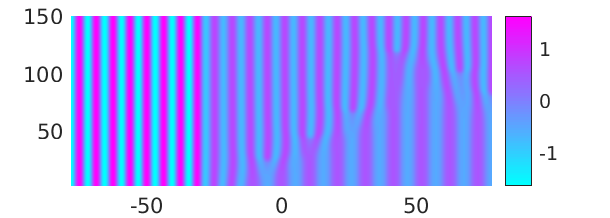}&
\ig[width=0.32\twi,height=20mm]{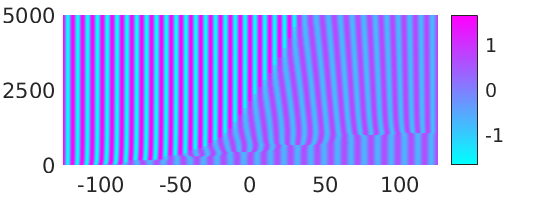}&
\ig[width=0.32\twi,height=20mm]{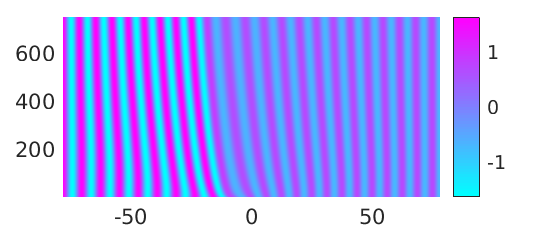}\\
(a2)&(c) &(e) \\
\ig[width=0.32\twi,height=22mm]{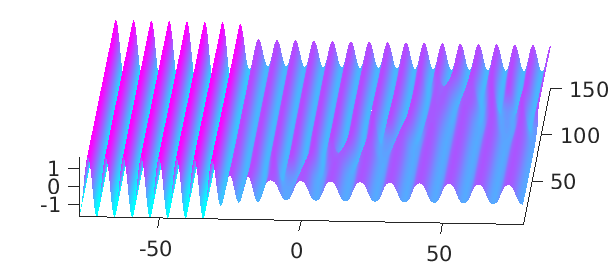}&
\ig[width=0.32\twi,height=20mm]{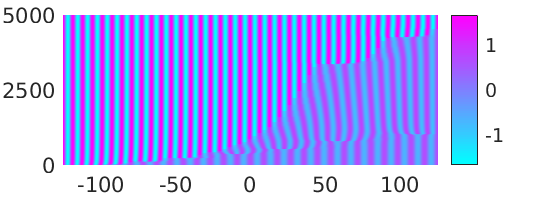}&
\ig[width=0.32\twi,height=20mm]{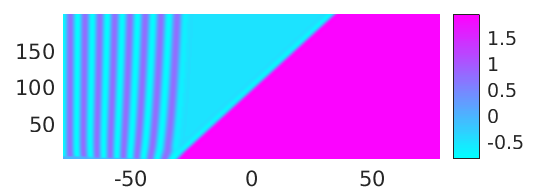}\\
\end{tabular}
\ece
\vs{-2mm}
\caption{{\small Time evolution on $\Om=(-25\pi,25\pi)$ (a,d,e) and $\Om=(-40\pi,40\pi)$ 
    (b,c) in the form of space-time diagrams. Horizontal axis shows space $x$, vertical axis
    time $t$, 
parameters as indicated.
Initial conditions $u_0$ consist of smooth amplitude and wavenumber transitions between one state on the left and a different state on the right (the precise form appears not to matter provided it is not too gradual). 
(a) $(a,\lam)=(5,1.3)$, 
$u_0=1.5\cos(x)$ for $x<-9\pi$, $u_0(x)=0.5\cos(0.6x)$ for $x\ge -5\pi$; (a2) shows 
the same solution as (a1) as a 3D plot. 
(b) $(a,\lam)=(5,1.49)$, $u_0=1.5\cos(x)$ for $x<-34\pi$, $u_0(x)=0.5\cos(0.6x)$ for $x\ge -28\pi$. (c) $(a,\lam)=(5,1.5)$, initial condition as in (b). 
(d) $(a,\lam)=(5,1.3)$, $u_0=1.5\cos(0.8x)$ for $x<-4\pi$, $u_0(x)=0.5\cos(x)$ for $x\ge 0$. 
(e)  $(a,\lam)=(9.5,3)$, $u_0=0$ for $x<-10\pi$, $u_0=2$ for $x\ge -10\pi$.
\label{f7}}}
\end{figure*}

The states satisfying \reff{Hd} correspond to time-independent structures
(fronts, pulses etc.) and the presence of snaking associated with these
connections implies that such structures remain stationary even away from the Maxwell
point, i.e., when $|\CE(u_1(\cdot;\lam)){-}\CE(u_2(\cdot;\lam))|{>}0$. If, however, the
energy difference between the two competing states becomes too large, the front
connecting them depins and the lower energy state invades the higher
energy state in a process analogous to depinning of fronts in SH23 or SH35
\cite{burke,burkeK}. Figure \ref{f7} shows examples of this process, 
focusing [panels (a-d)] on competition between states that are both spatially periodic, 
and illustrating possibilities that do not arise in either SH23 or SH35, namely 
front propagation via repeated phase slips. The fact that phase slips in SH357 can propagate is
of particular interest since the Eckhaus instability that triggers them is a steady state instability.
Since phase slips require a finite time for proceed to completion we may anticipate the presence of a
dynamic regime in which the speed of the front is determined by the phase-slip timescale and not the
energy difference alone \cite{MaK}. In (a), with $\lam=1.3$ in the snaking region 
of Fig.~\ref{f1}, these phase slips adjust the wavelength of the small amplitude 
portion, but the front between $\uplg$ and the $\upsg$ resulting from these phase slips remains
pinned. In (b) we have two fronts: a fast front consisting of propagating phase slips in $\upsg$,
followed by a slower front whereby the higher amplitude but lower energy state invades the smaller
amplitude higher energy state. The speed of this amplitude front is not constant, however, and is
strongly affected by the phase slips on the $\upsg$ portion. We conjecture that this is a
consequence of the fact that the $\upsg$ wave number has still not relaxed to its equilibrium value
when the amplitude front arrives.  In (c), corresponding to a slightly larger $\lam$ than in (b), the 
amplitude front triggers further phase slips, locally accelerating the front and resulting in a type
of stick-slip motion at large times. This does not happen in case (b) even at very long times.
Finally, in (d) the large amplitude part of the initial condition is dilated relative to its
equilibrium wavelength and in this case the system relaxes to a lower energy state via phase
diffusion in both $\upsg$ and $\uplg$, instead of phase slips. Here the front remains pinned
but its location adjusts accordingly.
 
Figure \ref{f7}(e) shows a multifront at larger $a$. Here we obtain a $\upsg$--$(-\uhs)$--$\uhl$
double front with retreating $\uhl$, obtained from a step-like initial condition, but 
as suggested by Fig.~\ref{f4}, all sorts of multifronts are possible via appropriate choice 
of initial conditions. Fronts between homogeneous states are then generically fast, while fronts
involving patterned states are typically substantially slower.

Equations of the 357 form have been considered before, in context of the Ginzburg-Landau
equation with real coefficients \cite{Bortolozzo}. When parametrically forced by a spatially
periodic function the homogeneous solutions of this equation become periodic solutions with
wavelength equal to the forcing wavelength. The resulting equation thus also exhibits
coexistence between different amplitude periodic states, and between periodic states
and the homogeneous state. While this equation also reveals the gradual breakup of
forced snaking into isolas \cite{Champneys} the situation is different since the wavelength
of the periodic states is imposed by the forcing wavelength with the result that bistability
between periodic states with different wavelengths is absent. Nevertheless models
of this type indicate that the phenomena described here may also occur
in periodically forced systems exhibiting periodic states with an intrinsic wavelength.
Such systems remain to be studied in detail, although preliminary studies of the spatially
forced Swift-Hohenberg equations SH23 and SH35 indicate that the snaking behavior is
likewise destroyed as the forcing amplitude increases \cite{Kao}.

In summary, the SH357 equation is an extremely rich pattern-forming system
and may be seen as a one-dimensional model for studying intricate structures in systems
exhibiting competition between states with distinct wavelengths. The gradient structure
of this equation, and in particular the existence of the spatial invariant $H$, allow
a more detailed understanding of the behavior of this equation than is possible for
other equations exhibiting similar behavior, such as the Gray-Scott model \cite{Gandhi2018},
but neither property is essential for the behavior described here as is well documented for
standard homoclinic snaking \cite{kno2015}. In particular, standard homoclinic snaking, as
described by SH23 or SH35, is robust with respect to both parameter changes and boundary
conditions \footnote{In the presence of nonperiodic or non-Neumann boundary conditions an extended
periodic state is absent and the snaking localized states turn continuously into a spatially extended
state satisfying the boundary conditions \cite{mbak2009}.}, and for these reasons is found in more
complex systems including the equations of hydrodynamics.

\noi
   {\bf Acknowledgment.} The work of EK was supported in part by the National Science Foundation under Grant No.~DMS--1613132. The work of DW was supported by the DFG under Grant No.~264671738.

\end{document}